\makeatletter
\declare@file@substitution{revtex4-1.cls}{revtex4-2.cls}
\makeatother

\documentclass[twocolumn]{aastex631}

\usepackage{float}      %
\usepackage{placeins}   %
\usepackage[hyphens]{url}
\usepackage{hyperref}   %

\shorttitle{R2Pub Telescopes for Surveying}
\shortauthors{Song et al.}

\begin{document}

\title{The R2Pub Telescopes for Surveying: An Overview and Performance Evaluation of the System }

\author[0009-0006-9242-1073]{\textnormal{SONG Xuan}}
\affiliation{Beijing Planetarium, Beijing Academy of Science and Technology, Beijing 100089, People’s Republic of China}

\author[0000-0002-7334-2357]{\textnormal{WANG Xiaofeng}}
\affiliation{Physics Department, Tsinghua University, Beijing, 100084, People’s Republic of China}

\author{\textnormal{ZHU Jin}}
\affiliation{Beijing Planetarium, Beijing Academy of Science and Technology, Beijing 100089, People’s Republic of China}

\author{\textnormal{LI Jian}}
\affiliation{Beijing Planetarium, Beijing Academy of Science and Technology, Beijing 100089, People’s Republic of China}

\author[0000-0002-8321-1676]{\textnormal{GUO Jincheng}}
\affiliation{Beijing Planetarium, Beijing Academy of Science and Technology, Beijing 100089, People’s Republic of China}

\author[0000-0002-1089-1519]{\textnormal{XIANG Danfeng}}
\affiliation{Beijing Planetarium, Beijing Academy of Science and Technology, Beijing 100089, People’s Republic of China}

\author[0000-0001-5879-8762]{\textnormal{LI Xin}}
\affiliation{Beijing Planetarium, Beijing Academy of Science and Technology, Beijing 100089, People’s Republic of China}

\author[0009-0005-4961-6070]{\textnormal{LIU Cheng}}
\affiliation{Beijing Planetarium, Beijing Academy of Science and Technology, Beijing 100089, People’s Republic of China}

\author[0000-0001-9442-1217]{\textnormal{NING Yuanhang}}
\affiliation{Beijing Planetarium, Beijing Academy of Science and Technology, Beijing 100089, People’s Republic of China}

\author[0000-0002-2614-5959]{\textnormal{GE Zhishuai}}
\affiliation{Beijing Planetarium, Beijing Academy of Science and Technology, Beijing 100089, People’s Republic of China}

\author[0000-0002-4448-3679]{\textnormal{SHAO Zhenzhen}}
\affiliation{Beijing Planetarium, Beijing Academy of Science and Technology, Beijing 100089, People’s Republic of China}

\author[0000-0002-0786-7307]{\textnormal{ZHENG Xiaochen}}
\affiliation{Beijing Planetarium, Beijing Academy of Science and Technology, Beijing 100089, People’s Republic of China}

\author{\textnormal{YANG Yi}}
\affiliation{Beijing Planetarium, Beijing Academy of Science and Technology, Beijing 100089, People’s Republic of China}

\author{\textnormal{ZHANG Lei}}
\affiliation{Beijing Planetarium, Beijing Academy of Science and Technology, Beijing 100089, People’s Republic of China}

\author[0009-0002-4524-3076]{\textnormal{SHI Yaqing}}
\affiliation{Beijing Planetarium, Beijing Academy of Science and Technology, Beijing 100089, People’s Republic of China}

\author[0000-0001-8592-7910]{\textnormal{ZHAO Dongyao}}
\affiliation{Beijing Planetarium, Beijing Academy of Science and Technology, Beijing 100089, People’s Republic of China}

\author[0000-0002-8049-202X]{\textnormal{ZENG Xiangyun}}
\affiliation{College of Mathematics and Physics, China Three Gorges University, Yichang 443000, People’s Republic of China}

\author{\textnormal{MO Jun}}
\affiliation{Physics Department, Tsinghua University, Beijing, 100084, People’s Republic of China}

\author[0000-0002-3876-6330]{\textnormal{SONG Tengfei}}
\affiliation{Yunnan Observatories, Chinese Academy of Sciences, Kunming 650216, People’s Republic of China}

\author[0000-0002-5210-2987]{\textnormal{FAN Yufeng}}
\affiliation{Yunnan Observatories, Chinese Academy of Sciences, Kunming 650216, People’s Republic of China}

\author[0000-0002-7694-2454]{\textnormal{LIU Yu}}
\affiliation{School of Physical Science and Technology, Southwest Jiaotong University, Chengdu 611756, People’s Republic of China}

\author[0009-0003-0581-4396]{\textnormal{WANG Jingxing}}
\affiliation{Yunnan Observatories, Chinese Academy of Sciences, Kunming 650216, People’s Republic of China}

\author{\textnormal{HE Shousheng}}
\affiliation{Yunnan Observatories, Chinese Academy of Sciences, Kunming 650216, People’s Republic of China}

\author{\textnormal{WANGDUI Ciren}}
\affiliation{Yunnan Observatories, Chinese Academy of Sciences, Kunming 650216, People’s Republic of China}

\author[0000-0002-8296-2590]{\textnormal{ZHANG Jujia}}
\affiliation{Yunnan Observatories, Chinese Academy of Sciences, Kunming 650216, People’s Republic of China}
\affiliation{International Centre of Supernovae, Yunnan Key Laboratory, Kunming 650216, China}

\author[0000-0002-3828-9837]{\textnormal{ZHANG Xuefei}}
\affiliation{Yunnan Observatories, Chinese Academy of Sciences, Kunming 650216, People’s Republic of China}

\author{\textnormal{YE Kai}}
\affiliation{Yunnan Observatories, Chinese Academy of Sciences, Kunming 650216, People’s Republic of China}

\author{\textnormal{BAI Jinming}}
\affiliation{Yunnan Observatories, Chinese Academy of Sciences, Kunming 650216, People’s Republic of China}

\author{\textnormal{JIANG Xiaojun}}
\affiliation{National Astronomical Observatories, Chinese Academy of Sciences, Beijing 100101, China}

\author{\textnormal{ZHANG Xiaoming}}
\affiliation{National Astronomical Observatories, Chinese Academy of Sciences, Beijing 100101, China}

\author{\textnormal{QIU Peng}}
\affiliation{National Astronomical Observatories, Chinese Academy of Sciences, Beijing 100101, China}

\author[0000-0002-4052-9932]{\textnormal{ZHANG Jicheng}}
\affiliation{School of Physics and Astronomy, Beijing Normal University, Beijing 100875, People’s Republic of China}

\correspondingauthor{WANG Xiaofeng}
\email{wang\_xf@mail.tsinghua.edu.cn}

\correspondingauthor{SONG Xuan}
\email{famcroo@gmail.com}

\begin{abstract}

The R2Pub, built by Beijing Planetarium, is a state-of-the-art 60\,cm equatorial binocular telescope located at the Daocheng Site (with an altitude of 4700\,m) of Yunnan Observatories in China. This paper provides an overview of the R2Pub telescope system, discusses its design and capabilities, and presents an evaluation of its performance for astronomical surveys. R2Pub is a prime-focus binocular system, with each tube covering a field of view of about 18 square degrees. This system is designed to detect various transients in local universe, including variables, eclipsing binaries, supernovae, gamma-ray bursts afterglow, tidal disruption events, Active Galactic Nuclei, and other unknown transients, which are ideal targets for both time-domain astronomy research and science population. The entire R2Pub system has completed the construction and installation of all observatory infrastructure, including the dome, equatorial mount, optical tube, and associated components, and has now entered the commissioning phase. The high-altitude location, good seeing, and dark background sky light at Daocheng site ensure optimal observational conditions for time-domain astronomy. Performance testing during the commissioning phase has demonstrated that the R2Pub system can achieve a 5$\sigma$ limiting magnitude of approximately 18.7 mag in the Pan-STARRS r' band for 60-second exposures. The ongoing observations from R2Pub is expected to contribute significantly to the study of time-variable phenomena in the universe and greatly improve the public outreach in astronomy. In particular, the binocular telescope systems capable of simultaneous dual-band observations can obtain the instantaneous color information of transient sources, enabling more accurate characterization of their physical properties and evolution, and providing a significant advantage in rapid distinguishing different classes of variables and transients.

\end{abstract}


\section{Introduction} \label{sec:intro}

\begin{deluxetable*}{lcccc}
  \tabletypesize{\small}
  \tablecaption{Key Parameters of Selected Time-Domain Survey Facilities\label{tab:surveys}}
  \tablehead{
  \colhead{Survey\tablenotemark{a}} &
  \colhead{\shortstack{Aperture (m)\\Tubes$\times$Mounts/Site\\Sites}} &
  \colhead{\shortstack{FoV\\(per tube / per mount)\\(deg$^2$)}} &
  \colhead{\shortstack{Limiting mag\\(filter, $t_{\rm exp}$)}} &
  \colhead{Cadence}
  }
  \startdata
  ASAS-SN       & 0.14 / 4$\times$1 / 2  & 4.5 / 18      & $\sim\,17 (V, 270\,s) $                    & $\sim$1$\times$/2–3 nights\\
  ATLAS(1-4)    & 0.5 / 1$\times$1 / 4   & 30 / 30       & $\sim\,19.7 (5\sigma, c/o, 30\,s) $        & 3–4$\times$/1 night \\
  ATLAS(5)      & 0.28 / 4$\times$4 / 1  & 7.3 / 7.3     & $\sim\,19–20 (c, 30\,s) $                  & same as ATLAS(1-4) \\
  BlackGEM      & 0.65 / 1$\times$3 / 1  & 2.7 / 8.2     & $\sim\,22 (5\sigma, $r$, 300\,s) $         & multi-visits/1 night (selected fields) \\
  GOTO          & 0.4 / 8$\times$2 / 2   & 5 / 44        & $\sim\,18.6 (5\sigma, $L$, 180\,s) $       & $\sim$1$\times$/2–3 nights \\
  LSST          & 8.4 / 1$\times$1 / 1   & 9.6 / 9.6     & $\sim\,24.5 (5\sigma, $r$, 30\,s) $        & $\sim$1$\times$/3 nights \\
  Mini-SiTian   & 0.3 / 3$\times$1 / 1   & 3.5 / 3.5     & $\sim\,19.5 (5\sigma,$g$, 300\,s) $        & $\sim$30 min revisit/field (gri bands)\\
  TMTS          & 0.4 / 4$\times$1 / 1   & 4.5 / 18      & $\sim\,19.4 (3\sigma, $L$, 60\,s) $        & multi-visits/night (selected fields) \\
  R2Pub         & 0.6 / 2$\times$1 / 1   & 18 / 18       & $\sim\,18.7 (5\sigma, $r$, 60\,s) $        & $\sim$1$\times$/2 nights \\
  TESS          & 0.1 / 4$\times$1 / 1   & 570 / 2300    & $\sim\,19.6 (3\sigma, W, 30\,min) $        & 27 d/sector \\
  WFST          & 2.5 / 1$\times$1 / 1   & 6.5 / 6.5     & $\sim\,23.0 (5\sigma, $r$, 30\,s) $        & see notes\tablenotemark{b} \\
  ZTF           & 1.22 / 1$\times$1 / 1  & 47 / 47       & $\sim\,20.6 (5\sigma, $r$, 30\,s) $        & $2$--$3\,\times$/1 night \\
  \enddata
  \tablecomments{FoV = field of view. Limiting magnitudes represent typical single-visit or coadded depths (filter, exposure). Cadence values are representative and strategy-dependent.}
  \tablenotetext{a}{
  ASAS-SN: \citet{Shappee2014, Kochanek2017};
  ATLAS(1-4): \citet{2011PASP..123...58T, 2018PASP..130f4505T};
  ATLAS(5): \citet{atlas2};
  BlackGEM: \citet{blackgem2024};
  GOTO: \citet{2018SPIE10704E..0CD, goto2};
  LSST: \citet{2009arXiv0912.0201L};
  Mini-SiTian: \citet{Han_2025, Liu2021AABC, minisitian3};
  TMTS: \citet{2020PASP..132l5001Z};
  TESS: \citet{tess};
  WFST: \citet{WFST2023, wfst2};
  ZTF: \citet{2019PASP..131a8002B}.
  }
  \tablenotetext{b}{
  WFST survey strategy: Wide Field Survey $\sim$20\,min revisits in multiple bands;
  Deep High-Cadence \textit{ugr} Survey $\sim$1\,hour revisits around new moon.
  }
\end{deluxetable*}

Time-domain astronomy is an emerging frontier of astrophysics that explores the temporal variations of celestial objects through repeated observations of the same region of the sky. The rapid development of this field has been driven by progresses in observational technologies, particularly breakthroughs in high-sensitivity detectors and high-time-resolution observational capabilities.

The continuous improvements of large-area CCD(charge--coupled device) detectors, CMOS(complementary metal--oxide--semiconductor
) detectors, and tiled large-aperture detector technologies, combined with advances in data processing and the introduction of robotic observation systems, have led to accelerating developments in time-domain astronomy. Over the past decade, several wide-field optical survey projects have emerged, significantly advanced the discovery and study of transient objects, such as supernovae (SNe), variable stars, active galactic nuclei (AGN), and tidal disruption events (TDEs). In recent years, there are many projects that have been proposed to detect and track these rapidly changing celestial objects in much shorter time frames using more sensitive and efficient technologies. 

One notable example is the Zwicky Transient Facility (ZTF; \citealt{2019PASP..131a8002B}) survey project. The ZTF project employs a large CCD detector array that offers a vast field of view of about 47 square degrees, with a pixel scale of 1.0 arcsecond per pixel, allowing for efficient coverage of the visible sky in the northern hemisphere in a short period of time. The typical observation cadence for the ZTF survey is 2–3 days, enabling timely monitoring and capturing the young phase of newly-discovered transients. 
Another well-known time-domain survey is the Asteroid Terrestrial-impact Last Alert System (ATLAS; \citealt{2011PASP..123...58T}), operating globally with a network of 0.5-meter telescopes, each covering approximately 30 square degrees with a resolution of 1.86 arcsecond per pixel. The ATLAS monitored the entire visible night sky roughly every 24 hours, providing exceptional cadence ideal for detecting rapidly evolving transients and potentially hazardous asteroids. 
Whereas, the All-Sky Automated Survey for SuperNovae 
(ASAS-SN; \citealt{Shappee2014}) is a global monitoring network using multi-tube 14-cm telescopes, with each having a FoV of about 4.5 square degrees.
ASAS-SN can monitor the entire visible sky every night down to $V \sim 17$~mag, providing uniform all-sky search of bright transients. 

Note that the Vera Rubin Observatory (LSST; \citealt{2009arXiv0912.0201L}), with an 8.4-meter aperture telescope and 3.2-billion-pixel CCD detector array, is becoming the most powerful time-domain project in the coming decades. It can view a 9.62 square-degree patch of sky with a detection depth down to 24.5 magnitude in a 30-second visit. 
It is expected to discover thousands of various transients every night, including numerous supernovae that can be used to probe dynamic evolution of the universe.

All of the above surveys are operated in single band and cannot obtain simultaneous color information for transients at the same time. The Tsinghua University--Ma Huateng Telescopes (TMTS; \citealt{2020PASP..132l5001Z}) is an array of four co-mounted tubes at Xinglong Observatory, with each having a FoV of about 4.5 square degrees. Although this telescope array can operate in two bands at the same time, it still operates in the $L$ band (Luminance $L$ filter in the LRGB system) over the past few years to maximize the monitoring sky areas. With relatively better angular resolution and deeper detection limit, TMTS discovered numerous short-period variable stars and eclipsing binaries \citep{lin2022, lin2023}. Owing to be located at the same site with LAMOST \citep{2012RAA....12.1197C}, it also provides great synergies with the spectroscopic observations.

Although the existing wide-field surveys have significantly advanced time-domain astronomy, 
synchronized multi-band data are still absent, which are crucial to place tight constraints on physical origins of some high-energetic transients, especially for SNe, kilonovas, TDEs and AGNs. For example, rapid color evolution in kilonovae \citep{1998ApJ...507L..59L,2010MNRAS.406.2650M}, early photometric behavior of Type~Ia SNe \citep{Nomoto1982a,Nomoto1982b,Woosley2011,Kasen2010,Lim2023}, and shock cooling signals in core-collapse SNe \citep{nature2024ixf,2013ApJ...767..143B,2017hsn..book..967W} all require near-simultaneous multi-band monitoring to constrain progenitors, explosion mechanisms, and circumstellar environments.

Based on the aforementioned objectives, we have constructed the R2Pub telescope system. The name R2Pub stands for ``from Research to Public outreach'', reflecting the dual mission of the Beijing Planetarium to focus not only on scientific research but also on science popularization. Fulfilling this mission also entails effective science communication, and active participation in various research areas of time-domain astronomy.
The R2Pub telescope is equipped with a dual-tube optical system, enabling simultaneous observations in two different photometric bands. This capability %
allows for precise color measurements, thereby improving the characterization of transients and variable sources. In this paper, we present a detailed description of the R2Pub system, including its hardware, software, and overall performance.

\section{Hardware System} \label{sec:Hardware}

\subsection{Enclosure} \label{subsec:Enclosure}

\begin{figure}[htbp]
  \centering
  \includegraphics[width=\linewidth]{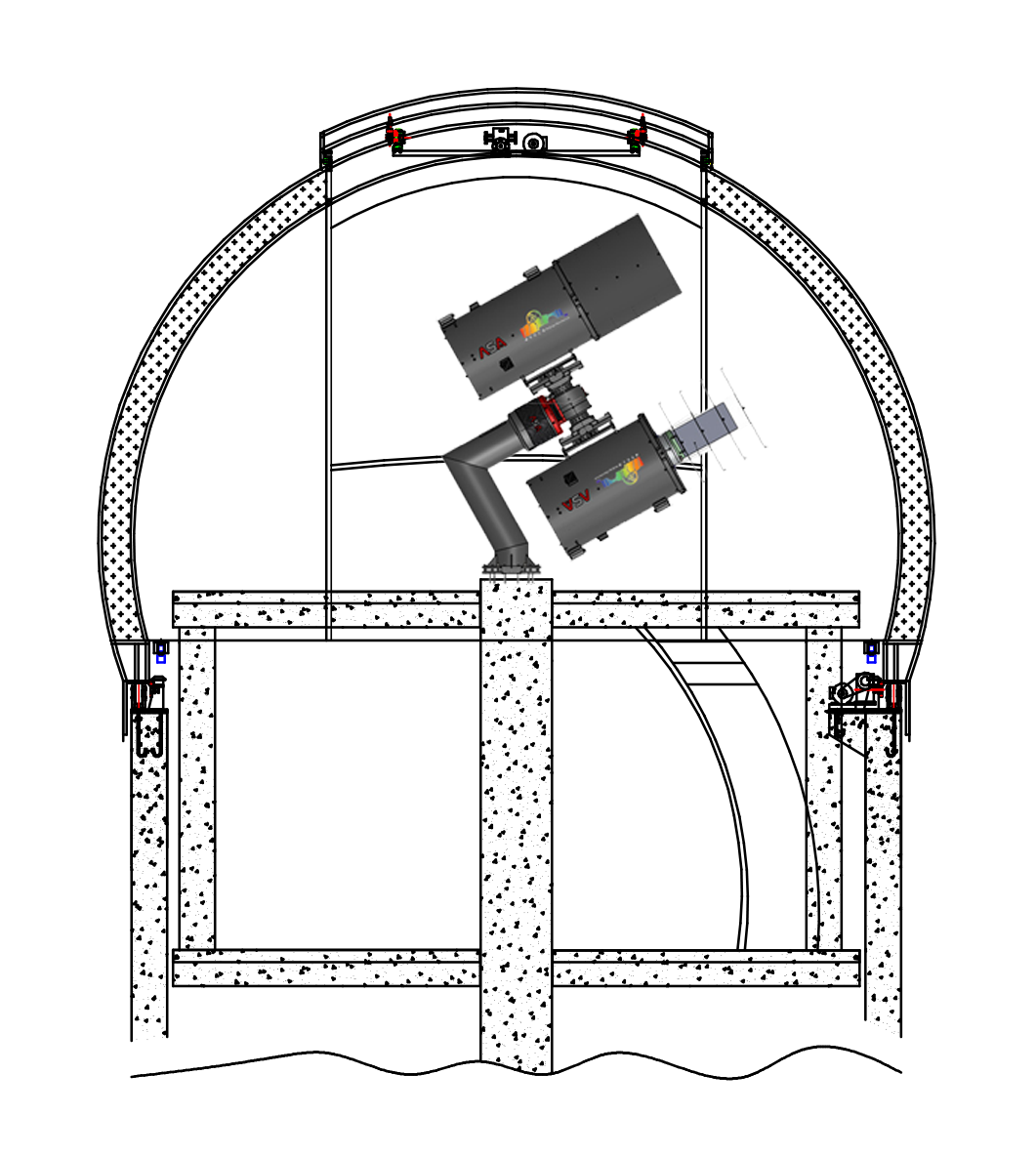}
  \caption{Layout of the shell of the R2Pub telescope. Adapted from Astro Systeme Austria GmbH (ASA), used with permission.}
  \label{fig:1}
\end{figure}

\begin{figure}[htbp]
     \centering
     \includegraphics[width=\linewidth]{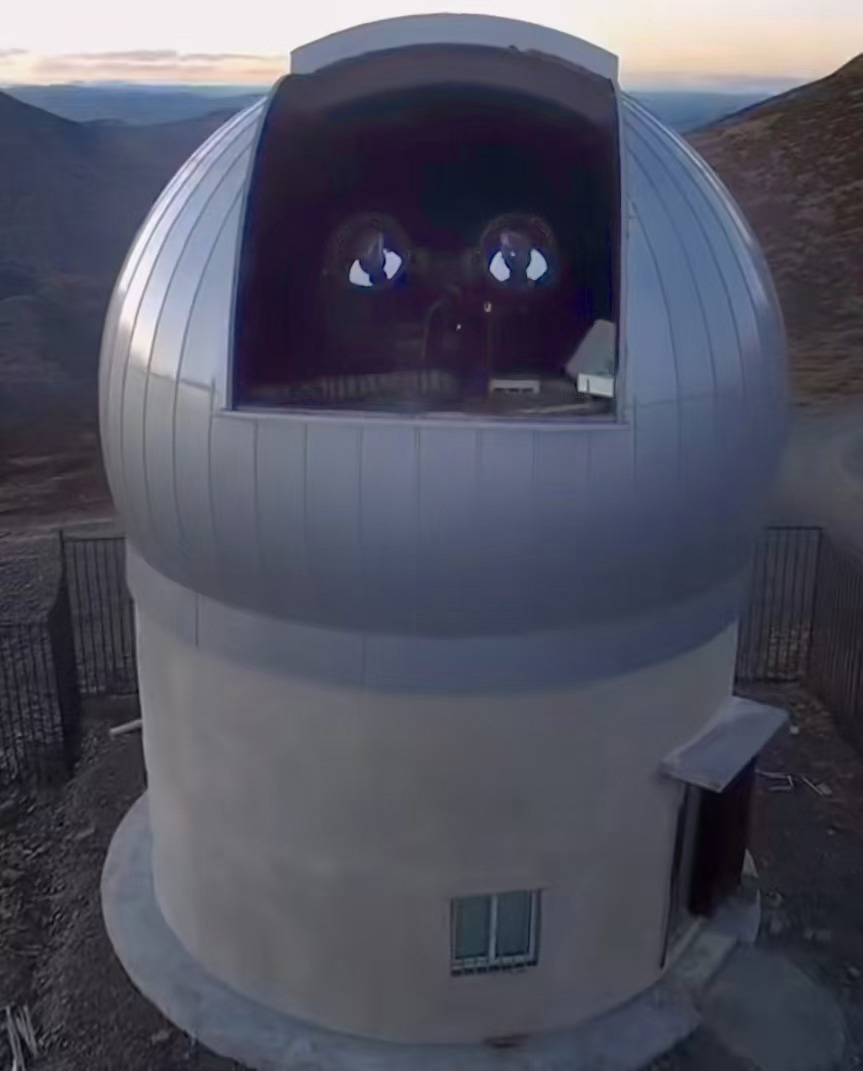}
     \caption{A real image of the R2Pub system installed inside the dome.}
     \label{fig:2}
\end{figure}

The dome of the R2Pub Observatory adopts a traditional observatory dome, which effectively reduces straylight interference. It can also enhance the wind load resistence of the system, and minimize the impact of humid air on internal equipments compared to the rolling roof or all-sky dome. To accommodate simultaneous observation with dual telescopes, the dome opening has been specially customized to a width of 3.1 meters. The design of the dome also ensures minimal obstruction at horizontal angles. Due to the excellent atmospheric transparency at Daocheng, 
we can set the elevation angle to be as low as 20 degrees. Figure \ref{fig:1} shows the layout of the R2Pub telescope system, while Figure \ref{fig:2} provides an overview of the R2Pub system inside the dome.

As shown in Figure \ref{fig:1}, the telescope's pier and the dome structure are supported independently. The central platform and telescope pier are integrated as a single unit, and are decoupled from the dome support system to minimize the vibration impact on the observations during the operation of the telescope. The control system of the dome is powered by a 380\,V supply and is equipped with a 380\,V UPS (uninterruptible power supply), backup power system. The control and data processing systems, however, use a 220\,V power supply, which is also supported by a separate UPS for backup.

The dome control system consists of three parts: rotation, opening/closing control, and position feedback. Power for dome rotation and opening/closing is supplied via a slip ring and guide rail, ensuring continuous power during the operation. Both dome rotation, opening/closing are driven by a gear-and-chain mechanism. To synchronize with the telescope's pointing, the dome rotation uses an incremental rotary encoder for position feedback and is equipped with a limit switch for precise positioning. The dome's actuation does not require continuous position feedback, but it is equipped with position-limited switches to ensure safety of the telescopes. It takes about 3 minutes to open or close the dome, and the dome can complete a full rotation in about 4 minutes. 

Regarding the control system, the dome uses an ASCOM(astronomy 
common object Model)-compatible driver developed based on the COM(component object model) components. The driver facilitates the integration of the dome and telescope control systems, and allows for further upgrades in the future. The ASCOM platform also supports remote operation, laying the foundation for remote control of the dome.
   
\subsection{Mount} \label{subsec:Mount}
Classical telescope mounts are generally categorized into two main types: alt-azimuth mounts and equatorial mounts. Both types have their respective advantages and disadvantages. Equatorial mounts can effectively avoid the issue of field-of-view (FoV) rotation and minimize the performance degradation near the zenith, making them more suitable for small-aperture telescope systems.

Each R2Pub tube weighs approximately 150\,kg, with an additional 21\,kg for the camera and extra weight from cables, cooling tubes, and other components, bringing the total weight to nearly 200\,kg. The DDM200 (DDM: direct--drive mount) equatorial mount, manufactured by Astro Systeme Austria (ASA), is adopted to support a payload of up to 200\,kg.

\begin{figure}[htbp]
  \centering
  \includegraphics[width=\linewidth]{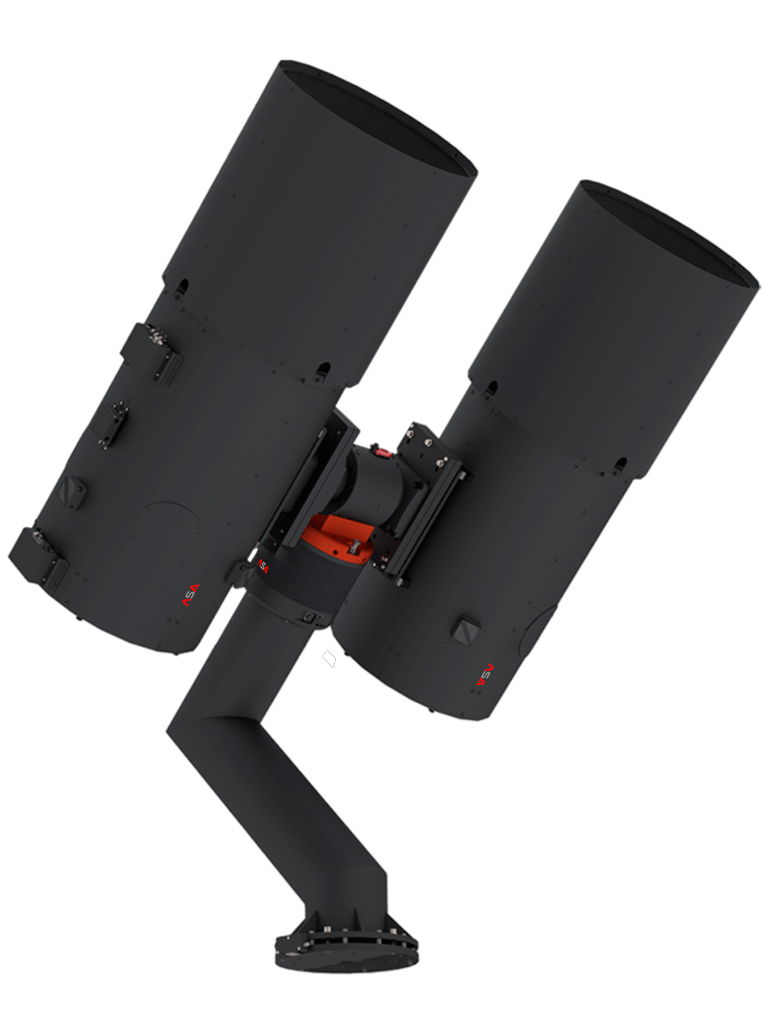}
  \caption{Illustration of the two tubes mounted on the DDM200 equatorial mount. Reproduced (with minor modifications) with permission from Astro Systeme Austria GmbH (ASA).}
  \label{fig:3}
\end{figure}

The configuration of the two telescopes on the DDM200 equatorial mount\footnote{\url{https://www.astrosysteme.com/en-us/products/asa-ddm200/}} is shown in Figure \ref{fig:3}.
Since both telescopes are mounted on the same platform, no additional counterweights are required, allowing the entire payload capacity to be utilized by the telescopes.
The specifications of the mount are listed in Table \ref{tab:mount_specs}.

\begin{table}[ht]
\caption{Specifications of the Mount ASA DDM200 }
\label{tab:mount_specs}
\centering
\begin{tabular}{lc}
\hline
\hline
Mount type & German equatorial mount \\
Maximum payload & 200\,kg \\
Maximum slew rate & $>10^\circ$ s$^{-1}$ \\
Latitude Range & $0^\circ$ to $90^\circ$ \\
Pointing accuracy & $<8''$ RMS with pointing model \\
Tracking accuracy & $<0.25''$ RMS in 5 minutes \\
\hline
\end{tabular}
\end{table}

The DDM200 mount utilizes industrial-grade direct-drive (DD) motors and high-resolution absolute encoders. The DDM200 equatorial mount utilized by the R2Pub system achieves a maximum slew speed of 10 deg s$^{-1}$. However, due to the facts that the dome takes about 4 minutes to complete a full rotation, high-speed tracking is not required for our scientific observations, and excessive acceleration may even be detrimental to the system, we thus limited the mount’s slew rate to 3~deg~s$^{-1}$. Even at this reduced rate, the telescope will frequently necessitate waiting for dome alignment.

When observing the sky areas ranging from $20^\circ$ to $85^\circ$ above the horizon, the DDM200 mount achieves an RMS pointing error of less than $8^{\prime\prime}$ without the need for guiding. Based on the pointing model, the tracking accuracy in 5 minutes is less than $0.25^{\prime\prime}$ RMS.

\subsection{Telescope \& Focuser} \label{subsec:TelescopeFocuser}

\begin{figure}[htbp]
  \centering
  \includegraphics[width=\linewidth]{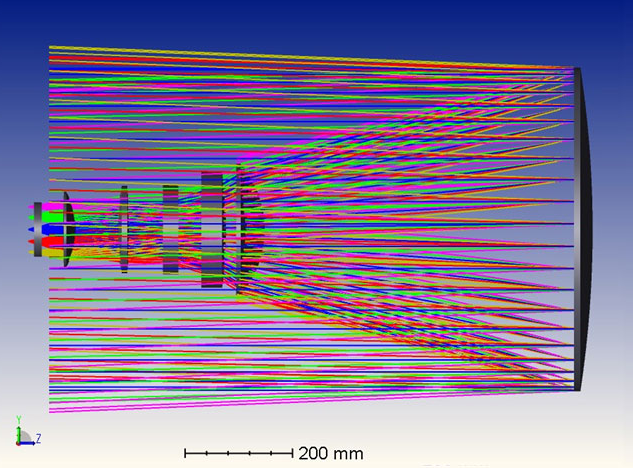}
  \caption{Illustration of the optical configuration of the R2Pub telescope. Reproduced with permission from Astro Systeme Austria GmbH (ASA).}
  \label{fig:5}
\end{figure}

Telescope performance is determined by key parameters including aperture, focal length, FoV, and image quality. 
The R2Pub system features two modified reflective prime-focus telescopes, which are custom-developed by ASA. 
Figure \ref{fig:5} shows the optical configuration of the R2Pub system, where the primary mirror is enclosed within a sealed tube, 
and a five-element corrector lens group positioned near the camera achieves optimized performance. 
Through this corrected design, the Ultra Wide Field 600 (UWF600) telescope\footnote{\url{https://www.astrosysteme.com/en-us/products/uwf600-f1-7/}} demonstrates significantly improved field flatness, 
reduced edge distortion, and stable imaging quality. 
The central obscuration ratio remains below 26\% even when accounting for combined obstructions from both the camera and corrector assembly. 
The sealed tube configuration effectively minimizes stray light interference and environmental impacts from high humidity.

The UWF600 telescope (f/1.7) employs three dedicated motors with corresponding absolute encoders at both the primary and corrector positions. This configuration enables not only compensation for misalignment of optical axis, focal plane tilt, and interface anomalies, but also achieves focal length adjustment through coordinated translation of optical components. Consequently, the system eliminates the need for a separate focusing.

\begin{table}[ht]
\caption{Specifications of the UWF600 Optical System}
\label{tab:UWF600_specs}
\centering
\begin{tabular}{lc}
\hline
\hline
Telescope diameter & 600 mm \\
Working f-ratio & $f/1.7$ \\
Focal length & 1020 mm \\
Central obscuration* & $<37.5\%$ \\
Back focus distance & 62.31 mm \\
Image field & 84 mm \\
System FoV (with R2Pub camera) & $\sim 18$ deg$^{2}$ \\
Encircled energy & 85\% within 0\farcs6 \\
Designed working wavelength & 400 nm $\sim$ 900 nm \\
Mirror material & \shortstack{Russian K8 \\ (like Schott BK-7)} \\
\shortstack{Diffraction limited surface quality \\ (P--V wave front)} & Minimum of 1/4$\lambda$ \\
Mirror coating reflectivity & 96\% \\
Designed rms spotsize & 1.76 $\sim$ 3.46 $\mu$m \\
OTA weight & $\sim$50 kg \\
\hline
\end{tabular}
\end{table}

Table \ref{tab:UWF600_specs} presents the specifications of the modified reflective prime-focus design. 
Each telescope is equipped with a 600~mm diameter primary mirror fabricated from high-precision optical substrate, delivering a system Strehl ratio greater than 94\%. 
With a primary focal length of 1020~mm and a fast focal ratio of f/1.7, 
the optical system operates effectively across the wavelength range of 400--900~nm. 
This combination of aperture, focal ratio, and broadband performance facilitates excellent image quality with reduced exposure times, 
making the system ideal for high-throughput astronomical surveys and time-domain observations.

The optimized corrector delivers an 84~mm diameter corrected image circle at the focal plane, yielding a wide and well-corrected FoV. With the camera configuration adopted for the R2Pub system, this optical design provides a system-level field of view of approximately 18~deg$^{2}$, while the detailed FoV calculation is presented in section 2.5. This optical configuration ensures high resolution with minimal distortion across the usable field, making it specifically tailored for wide-field surveys and precision astronomical observations.

\subsection{Filter} \label{subsec:Filter}

Considering the effective FoV of the R2Pub system and the position of the filters in the optical path, we customized circular filters with a diameter of 140\,mm and a thickness of 3\,mm. Since the telescope uses a prime focus design and does not allow for the installation of an automatic filter wheel, we designed a manual filter drawer for filter replacement.

\begin{figure}[htbp]
  \centering
  \includegraphics[width=\linewidth]{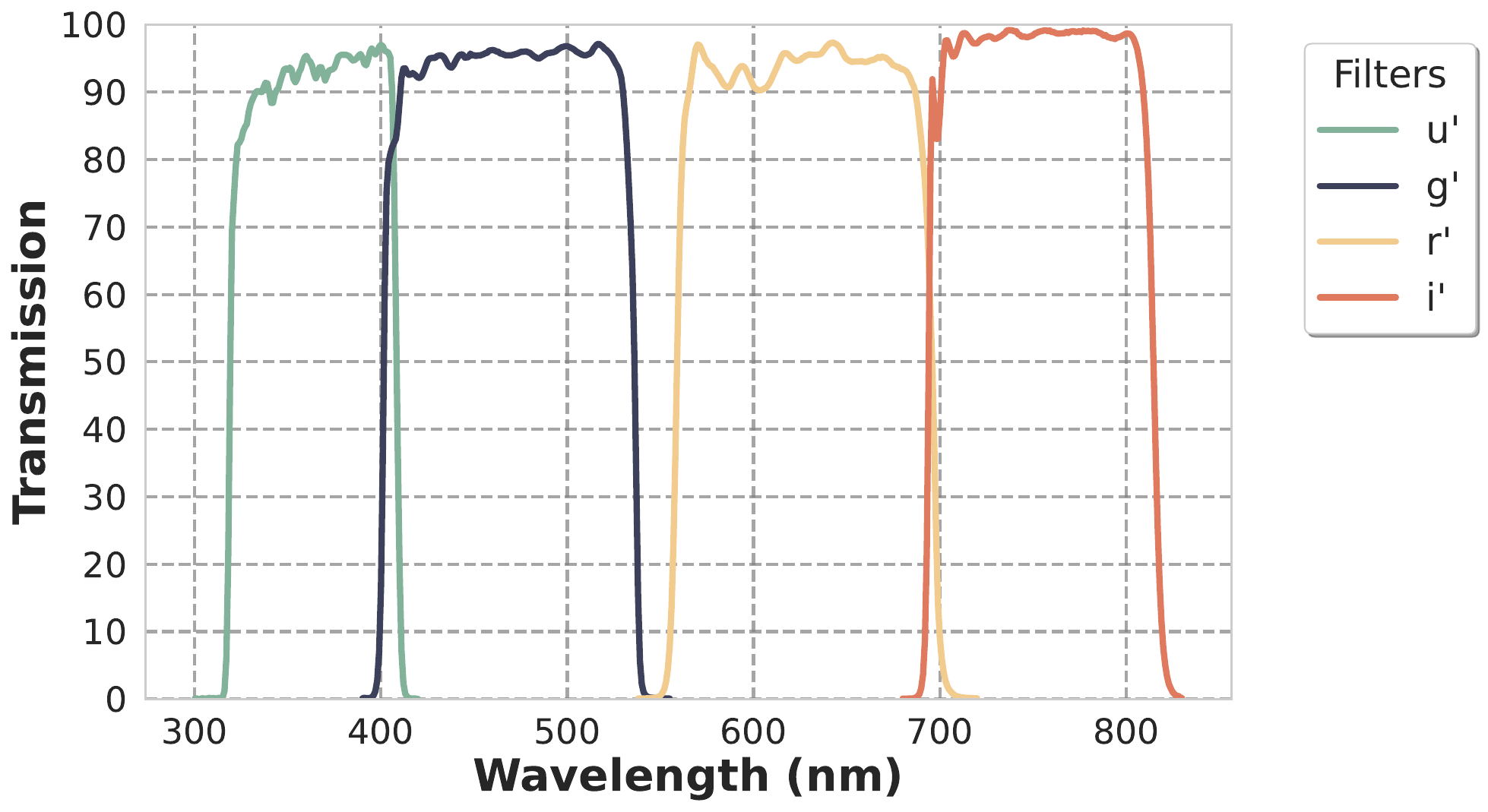}  %
  \caption{The transmission Curves of the R2Pub filters in \( u' \), \( g' \), \( r' \), and \( i' \) bands.}
  \label{fig:filters}
\end{figure}

Based on the SDSS filters and the transmission characteristics of the telescope itself, we customized four filters: \( u' \), \( g' \), \( r' \), and \( i' \). Their central wavelengths are 3600 \(\text{\AA}\), 4700 \(\text{\AA}\), 6250 \(\text{\AA}\), and 7550 \(\text{\AA}\), respectively, with corresponding full width at half maximum (FWHM) values of 900 \(\text{\AA}\), 1350 \(\text{\AA}\), 1380 \(\text{\AA}\), and 1210 \(\text{\AA}\). Figure \ref{fig:filters} displays the transmission curves of the R2Pub system's filters.

For the first phase of the survey, we plan to use the \( g' \) and \( r' \) filters in the two telescopes. This combination will allow us to simultaneously capture the \( g' - r' \) colors for all survey targets. In subsequent survey phases, we will experiment with combinations of filters such as \( u' \) and \( i' \).

\subsection{Camera} \label{subsec:camera}

\begin{table}[ht]
\caption{Specifications of the COSMOS-66 Camera}
\label{tab:COSMOS6k_specs}
\centering
\begin{tabular}{lc} 
\hline
\hline
Model & PI COSMOS-66 \\
Effective pixel area & 8120 x 8120 \\
Pixel Size & 10\,$\mu m \times 10\,\mu m$ \\
Sensor Area & 81.2\,mm x 81.2\,mm \\
Peak QE\% & $> 86\%$ peak QE* \\ 
Full Well Capacity & ~ 14k\ e- (typical, high gain) \\
Readout Noise & $< 1.5$ e- RMS (high gain mode) \\
Readout Modes & Rolling and Global shutter \\
Dark Current & $< 0.05$ e-/p/s (typical) \\
Cooling Method & TEC with liquid circulation  \\
Cooling Temperature & $< -{25}^\circ\mathrm{C}$ (guaranteed) \\
Bit Depth & 14-, 16- \\
Window Material & UV grade quartz glass** \\
Camera Weight & $~ 21.4\,kg$ \\
Nonlinearity & $< 1\%$ \\
\hline
\end{tabular}
\tablecomments{*Related to coating, see the plot of QE curve. \\**The camera window is made of UV-grade fused silica, specified by the manufacturer as JGS1.}
\end{table}

As part of our sky-survey program, we selected the PI COSMOS-66 \footnote{\url{https://www.teledynevisionsolutions.com/products/cosmos/}} to maximize the FoV, a large-format CMOS camera whose performance and characteristics have been extensively characterized by \citet{2025JATIS..11b6003L}. With its $81.2\,\mathrm{mm}\times81.2\,\mathrm{mm}$ active area and the UWF600 telescope of focal length $f=1020\,\mathrm{mm}$, the geometric field along each axis is $4.56^\circ$, i.e., about $20.8\,\mathrm{deg}^2$ for the full-frame. The sensor diagonal is $114.8\,\mathrm{mm}$, which exceeds the UWF600’s nominal $84\,\mathrm{mm}$ corrected image circle. To avoid degraded edge performance, we conservatively adopt a usable field of $\sim18\,\mathrm{deg}^2$, and use this value throughout the paper.

The PI COSMOS-66 detector is a single-chip CMOS with a resolution of 8120 $\times$ 8120 pixels, each pixel measuring 10\,$\mu m \times 10\,\mu m$. The pixel size, combined with the resolution of the UWF600 telescope, results in a resolution of 2.02$^{\prime\prime}$ per pixel. The median value of the seeing quality at Wuming Mountain Observatory of Daocheng, where the R2Pub telescope is located, is 0.99$^{\prime\prime}$ \citep{songtengfei2020, liuyu2018}, which is monitored in an under-sampled situation. However, given the primary goal of maximizing survey efficiency and considering the factors like focal length changes and vibrations caused by wind, the 2.02$^{\prime\prime}$ per pixel resolution is acceptable for time-domain observations.

The UWF600 has a back focus from the flange of 62.31\,mm, which can support a total distance of 24.1~mm from the sealing gasket of the PI COSMOS-66 detector to the sensor location. Using a custom-designed lightweight adapter, we connected the detector to the telescope while also incorporating a filter drawer into the adapter.

By utilizing the PI COSMOS camera SDK, we developed an asynchronous imaging program, allowing the camera to capture and save images concurrently. While one image is being captured, the previous image data is saved. The size of each individual image from the PI COSMOS-66 is approximately 127\,MB, and the readout and saving time per frame is well under 1 second. Our survey cadence is normally 10 seconds, with a shortest cadence of 1 second. This setting ensures the total readout and saving time acceptable.

The PI COSMOS-66 camera currently provides two primary readout configurations: a 14-bit rolling shutter mode and a 16-bit global shutter mode. In the rolling shutter configuration, the high-gain readout exhibits a noise level of 1.4\,e$^{-}$, whereas the 16-bit global shutter mode shows a higher readout noise of $\sim$6\,e$^{-}$. For our standard 10\,s exposures, the 14-bit rolling shutter mode is adopted, as it yields a superior signal-to-noise ratio under these conditions. For 10\,s exposures, the distinction between rolling and global shutter is negligible. Thus, the noise level is the decisive factor. Additional operating modes, such as an 18-bit readout and a 16-bit high-CMS(correlated multi--sampling) 8$\times$ mode, are included in the hardware design but are not supported in the current firmware. These advanced modes will be evaluated following future firmware upgrades.

The PI COSMOS camera system employs a liquid cooling solution utilizing two 1000\,W compressor-driven water chillers with pure ethylene glycol coolant. However, this configuration encountered operational challenges when ambient temperatures dropped below 3$^\circ\mathrm{C}$, leading to compressor startup failures. This low-temperature limitation issue was successfully solved by implementing environmental heating measures around the chillers' operational area to maintain adequate thermal conditions for proper system initialization.

\subsection{Network and Computing Infrastructure}\label{subsec:netAndComputer}
\begin{figure}[htbp]
  \centering
  \includegraphics[width=\linewidth]{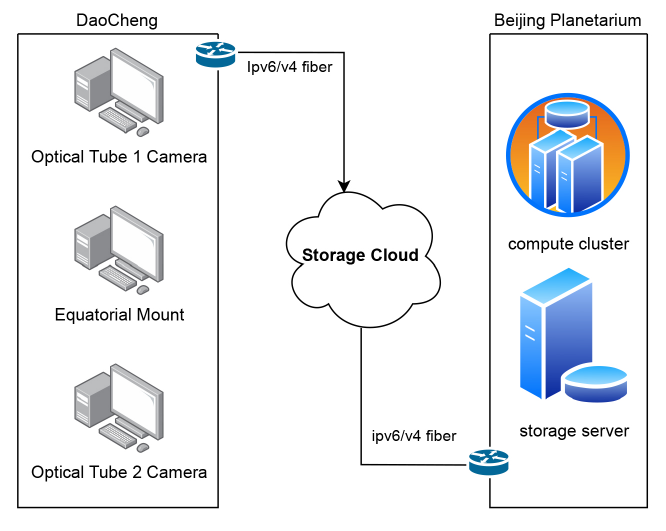}  %
  \caption{Network and Computing Infrastructure}
  \label{fig:networkcomp}
\end{figure}
Located in Daocheng, Sichuan province, at an altitude of 4,700 meters, the R2Pub telescope operates as an unmanned facility under extreme climatic conditions. To meet stringent reliability requirements for computer systems, we deployed three industrial control computers. Two control dual detectors for imaging operations, while the rest one manages the equatorial mount and dome. 

The observatory leverages a fiber-optic internet connection for data transmission. Due to the network instability when connecting directly from Daocheng to the Beijing Planetarium, we employ cloud storage as an intermediary to solve the data synchronization issue. 
With this approach, high reliability is achieved with only moderately decreased synchronization speed. Images are directly uploaded to the cloud via this link, with each frame taking approximately 3--4 seconds to transfer. At the same time, data processing workstations and redundant storage systems at the Beijing Planetarium automatically download these cloud-synced images at a rate of about 10 seconds per file. In the future, we plan to deploy localized data processing workstations at Daocheng Observatory to enable real-time processing of sky survey data.

\section{Software System and Survey Strategy} \label{sec:software}

\subsection{Software System} \label{subsec:soft}
The R2Pub software system is structured by two core subsystems: the Observation Subsystem and the Data Processing Subsystem. The Observation Subsystem accepts inputs such as observation schedules or General Coordinates Network (GCN) alerts. Then it commands the telescope to acquire exposures of specified targets. The Data Processing Subsystem performs preprocessing of raw images, including template subtraction, photometry, and transient detection. The Observation Subsystem runs primarily on a Windows platform, while the Data Processing Subsystem operates mainly on Linux system.

\subsubsection{Observation Subsystem} \label{subsubsec:obssoft}
The observation subsystem comprises several dedicated modules: telescope tube control, equatorial mount control, camera control, dome control, weather warning and monitoring, and a central scheduling system. Each module operates with its own specialized control software, integrated via \texttt{Python} scripting, and is coordinated under the command of the central control module. These modules are described below.

\textbf{Camera Control Module}. Each R2Pub telescope is equipped with an individual camera. Therefore, simultaneous exposure is required during the observations. To accomplish this, the camera control subsystem employs the general Remote Procedure Call (\texttt{gRPC}) framework in \texttt{Python}, which is structured in a client-server architecture. Each camera connects directly to its own server, simultaneously communicates with a shared client for control. 

On the server side, the Teledyne-provided SDK is integrated through \texttt{Python} to manage camera operations, including cooling, exposure triggering, data readout, and image storage. Detailed camera parameters are embedded and preserved within the FITS headers. Image data are saved using lossless \texttt{RICE\_1} compression, significantly reducing file sizes for efficient data transfer and storage. A flat-field imaging module is also implemented, which dynamically estimates appropriate exposure times based on real-time flux measurements toenable adaptive calibration under varying sky conditions. 

The \texttt{gRPC} interface enables efficient non-blocking communication. It facilitates simultaneous exposure across multiple cameras and seamless bidirectional parameter exchange between servers and the central clients.

The current software-based synchronization over \texttt{gRPC}, with each camera connected through a CoaXPress data interface, achieves a relative timing precision of approximately 1 ms, limited mainly by network and system scheduling latency. COSMOS-66  supports deterministic hardware triggering with microsecond or sub-microsecond latency and jitter under ideal conditions. A planned upgrade will therefore implement GPS-disciplined hardware triggering to realize this level of synchronization accuracy, with absolute timing traceable to UTC.

\textbf{Equatorial Mount Control}. 
Telescope pointing, tracking, and parking operations are performed using \texttt{Autoslew} software, which is featured by an intuitive graphical user interface (GUI). \texttt{Autoslew} also supports the ASCOM interface, allowing the central control module to seamlessly access real-time mount positional data and operational control, thereby ensuring precise and automated telescope guidance.

\textbf{Dome Control}. 
Dome operations, including rotation, tracking synchronization with the telescope, and shutter mechanisms, are managed via both a GUI and the ASCOM interface. This allows operators and scripts to control the dome in a flexible manner, ensuring efficient and reliable dome operation.

\textbf{Telescope Tube Control}. 
The telescope tube control is specialized, consisting of three primary mirror motors and three corrector mirror motors. Each motor has an individual distance sensor andis independently manageable via the ASA Gateway GUI. This GUI additionally provides synchronized control of the primary mirror motors in order to achieve precise focus adjustments. 

\textbf{Weather Warning and Monitoring}. 
Environmental conditions—including temperature, humidity, wind speed, and wind direction—are continuously monitored. The central control module uses these data to assess observing feasibility, automatically triggering dome closures in response to adverse weather conditions. Furthermore, it monitors UPS and other critical infrastructure to guarantee timely dome closure and equipment safety in power interruptions.

\textbf{Central Control Module}. 
The central control module integrates functionalities via \texttt{gRPC} and ASCOM interfaces, serving as the operational backbone of the subsystem. Its internal scheduling system ingests an observation list and automatically coordinates telescope pointing and exposure operations. This versatile scheduling supports continuous monitoring of single sky regions or expansive sky scanning tasks. 

The scheduling system handles the following tasks:

\begin{itemize}
  \item \textbf{Meridian Flip:} When the observation target approaches the meridian, the scheduling system automatically performs a meridian flip to ensure the telescope can continue observing as it moves toward the western sky.
  
  \item \textbf{Avoiding Southwest Winds:} Due to the prevalent southwest winds at the Daocheng site, the scheduling system avoids pointing the telescope to the southwest when wind speeds exceed level 4, which can be measured from weather probes. The system will automatically switch to the next observation target to prevent vibrations that could affect image quality.
  
  \item \textbf{Scheduled and Emergency Dome Closure:}  Scheduled dome closures upon completing all observational tasks and emergency closures triggered by critical events such as power outages, network disruptions, or deteriorating weather conditions are supported. 
  
  \item \textbf{Safety Angle Limits:} The \texttt{Python} package \texttt{Skyfield} is used to calculate safety angular distances to the Moon and other potential hazards, as well as safety altitudes. If an observation target is close to these limits, the system will check whether the target is within safety parameters before proceeding. If the target exceeds safety limits, the scheduling system will switch to the next target.
\end{itemize}

The central control's observation scheduling relies on a specialized list-generation tool developed in \texttt{Django}. This tool facilitates collaborative integration of diverse observational requests across varying targets and timelines, substantially enhancing telescope utilization efficiency and reducing observer's workload.

\subsubsection{Data Reduction Subsystem} \label{subsubsec:datareduction}

The R2Pub data reduction subsystem is tailored according to scientific goals. For short-period variable sources, the primary objective is to extract minute-scale light curves. We developed and enhanced processing code based on projects such as \texttt{STDPipe} \citep{stdpipe} and \texttt{AutoPhOT} \citep{Brennan_2022}, establishing an efficient pipeline for photometry of short-period variables. The data processing workflow consists of the following steps: 

\textbf{1. Data Reduction:} Raw images undergo bias subtraction, flat-field correction, and stacking. Initially implemented using \texttt{CCDProc}, the process was optimized by rewriting key parts in \texttt{CuPy}, utilizing GPU acceleration to handle the large (8120×8120 pixel) data efficiently. 

\textbf{2. Initial WCS Registration:} Original frames lack complete WCS information. Thus, equatorial mount RA/Dec data preserved in FITS headers facilitates WCS solutions.  \texttt{ASTAP} is adopted due to its speed and versatility in handling varying FoV. Those unresolved images are discarded as defective. 

\textbf{3. Source Extraction:} Precise RA/Dec positions of all sources are obtained using \texttt{SExtractor}. This step includes bad pixel and cosmic-ray marking, followed by extraction using a default aperture based on an FWHM of 3.5.

\textbf{4. Precise WCS Registration:} Utilizing initial WCS data and extracted sources, \texttt{SCAMP} refines WCS solutions using the Gaia DR3 catalog as a reference. 

\textbf{5. Photometry:} Background-subtracted detailed photometry is conducted to measure instrumental magnitudes, magnitude errors, and precise positions using \texttt{SExtractor}. Columns such as MAG\_BEST and MAG\_AUTO are retained, while aperture photometry results are saved in MAG\_APER, using the median FWHM from the initial measurement. 

\textbf{6. Flux Calibration:} Instrumental magnitudes are converted into AB magnitudes using reference stars from the Pan-STARRS catalog. Sources are quality-controlled based on Pan-STARRS magnitudes and MAGERR\_AUTO criteria, followed by weighted linear fitting with Statsmodels' WLS method for final AB magnitude calculation. 

Subsequently, deeper images are generated by combining multiple calibrated frames using \texttt{SWarp}. Template images are created for specific sky regions. Applying image subtraction with Hotpants, transient sources are then identified in the template-subtracted images.

\subsection{Sky Coverage and Survey Strategy} \label{subsec:skycoverage}
\begin{figure}[htbp]
  \centering
  \includegraphics[width=\linewidth]{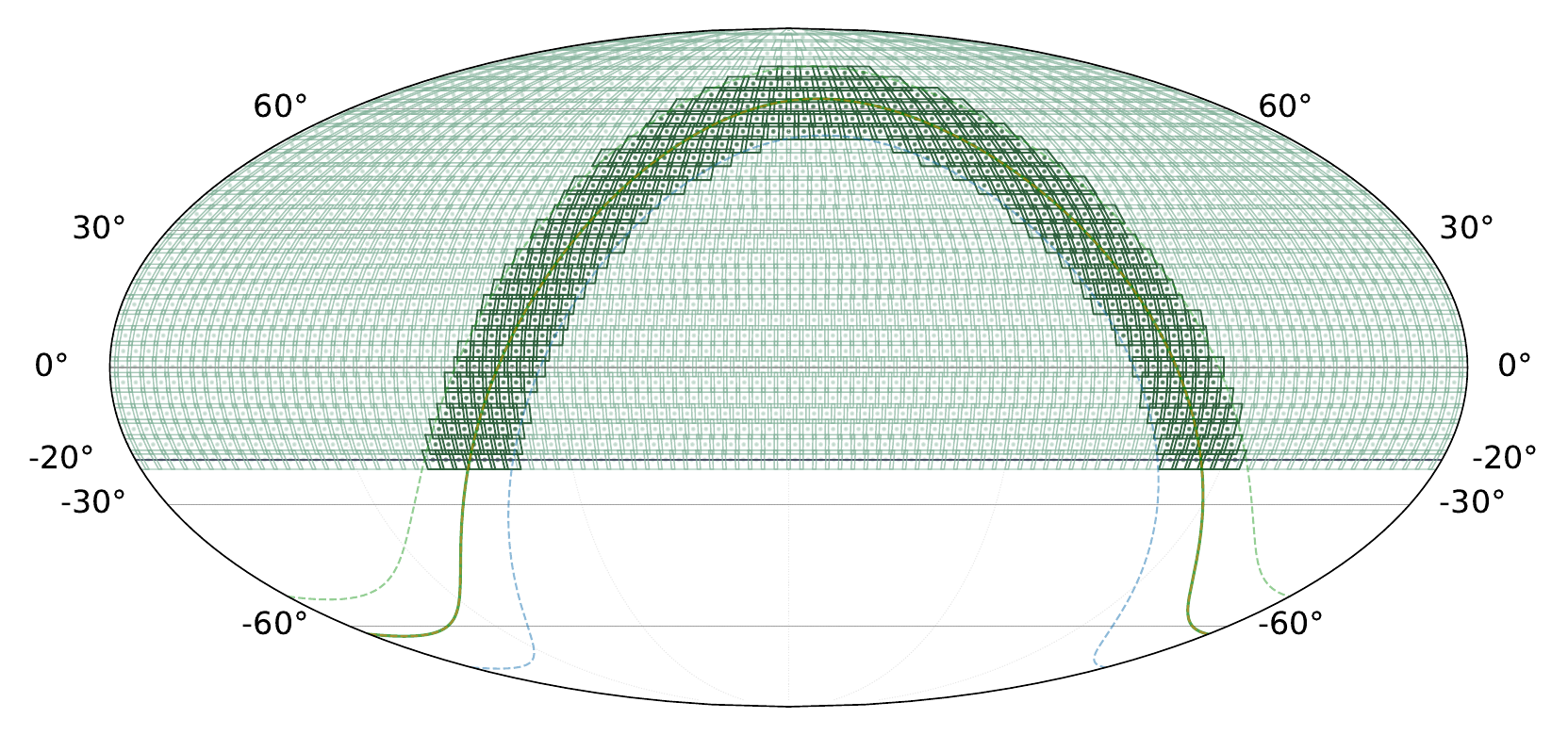}  %
  \caption{Sky coverage of the R2Pub survey. The green grid represents the extragalactic fields targeted by the supernova survey, covering declinations from $-20^\circ$ to $+90^\circ$. The darker band along the Galactic plane corresponds to the short-period variable survey, defined as regions with Galactic latitude $|b|<10^\circ$. Each field covers $\sim18\,\mathrm{deg}^2$ with a 20\% overlap between adjacent frames.}
  \label{fig:skycoverage}
\end{figure}
Each frame of the R2Pub system covers approximately $18\,\mathrm{deg}^2$, enabling the survey to span declinations from $-20^\circ$ to $+90^\circ$, thereby reaching the north celestial pole. Figure~\ref{fig:skycoverage} shows the distribution of survey fields, each covering $\sim18\,\mathrm{deg}^2$ with simultaneous observations in $g'$ and $r'$ bands. To mitigate edge effects and avoid placing transient sources on the borders of adjacent frames, a field overlap factor of 0.2 is adopted, corresponding to a 20\,\% overlap between neighboring fields. Here, an overlap factor of 0 denotes no overlap, while a value of 0.5 represents a 50\,\% overlap.

Our sky survey employs two distinct strategies. The first is the transient survey, which primarily targets extragalactic regions outside the Galactic plane (light green areas in Figure~\ref{fig:skycoverage}). The second is the short-period variable survey, focusing on the Galactic plane, defined here as regions with Galactic latitude $|b| < 10^\circ$ (dark green areas along the plane in Figure~\ref{fig:skycoverage}). 

For the short-period variable survey, each field is observed continuously for 2–3 hours or more cycles with individual exposure of 10~s, allowing accurate light-curve measurements of rapidly varying stellar sources. In the supernova survey mode, each targeted field receives a cumulative exposure of either 60~s or 90~s, achieved through consecutive 10~s frames to avoid detector saturation. During data processing, these consecutive frames are combined into a single stacked image, which facilitates the removal of cosmic rays and satellite trails. The daily survey schedule is generated automatically, prioritizing fields near the meridian to minimize slewing overheads and ensuring a suitable revisit cadence.

Operating the two telescopes in a deployed configuration can effectively double the single-frame FoV. This mode, however, precludes simultaneous acquisition in both photometric bands. The observing modes can be dynamically reconfigured using the declination control mechanism of the equatorial mount.

At a 60-second cumulative exposure, the survey covers approximately 50 sky regions per hour. During the winter observational season at 
Daocheng, the dual-band mode facilitates a coverage of approximately 7,000 square degrees per night.

\section{R2pub Performance and Initial Scientific Results} \label{sec:r2pub_performance_and_results}
In this section, we discuss the performance of the R2Pub system based on a pilot survey conducted in recent months. Observations were carried out using both the $g'$ and $r'$ filters. The resulting light curves are compared with data from ZTF. Preliminary comparison results are presented here, while detailed scientific analysis and findings will be reported in future publications.

\subsection{Mount Performance} \label{subsec:mount_performance}

Upon completion of the installation, we constructed a pointing model for the R2Pub system. Although ASA's proprietary sequence software is capable of automatically selecting target points, performing imaging, and executing plate-solving to generate pointing models, it relies on \texttt{Maxim DL} software. However, the \texttt{Maxim DL} is incompatible with our camera due to the substantial size of single-frame data. Additionally, our camera and mount are operated on separate control computers, further precluding the use of sequence. To address these constraints, we developed a suite of custom software tools. These include a point-selection algorithm based on uniformly distributed points on a sphere generated from a Fibonacci sphere method, an imaging tool that captures images sequentially by azimuth coordinates, and a positional error calculation tool using \texttt{ASTAP} software. These computed offsets are subsequently fed into the mount’s control software \texttt{Autoslew}, to build the pointing model.

\begin{figure}[htbp]
  \centering
  \includegraphics[width=\linewidth]{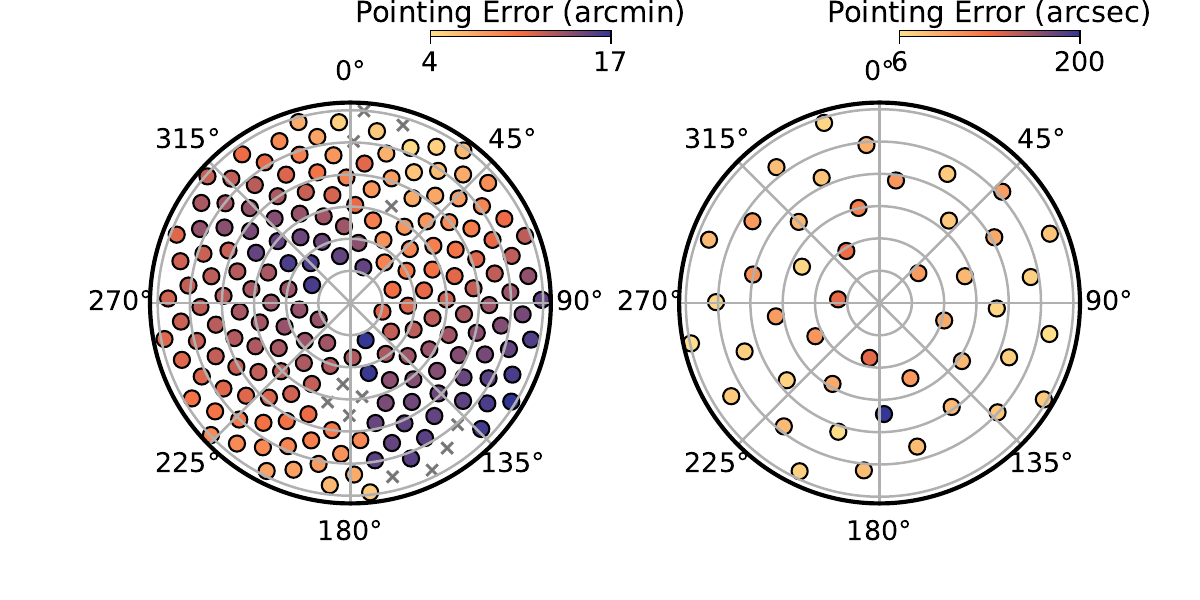 }  
\caption{%
\textit{Left}: The positions selected for building the pointing model of R2Pub. The panel shows 196 uniformly-distributed target positions atelevations between $30^\circ$ and $80^\circ$, color-coded by the magnitude of pointing errors before model correction. Positions missed due to cloud interference are marked with crosses.
The final dataset yielded 182 valid points, with an RMS error of $726.54^{\prime\prime}$ and a maximum error of $1019.93^{\prime\prime}$.
\textit{Right}: Pointing error distribution after applying the pointing model, where the RMS error was reduced to $57.01^{\prime\prime}$, the median error to $36.91^{\prime\prime}$, and the maximum error to $198.34^{\prime\prime}$.}

  \label{fig:pointing_compare}
\end{figure}

The left plot of Figure \ref{fig:pointing_compare} shows the selected pointing positions, consisting of 196 evenly distributed points at elevations between 30° and 80°. Regions near the northern celestial pole are intentionally avoided. Colors denote the magnitude of pointing errors before applying the pointing model. Due to cloud interference, some positions were omitted and are marked with crosses. The final dataset contains 182 valid data points, yielding an RMS pointing error of 726.54 arcsec and a maximum error of 1019.93 arcsec. The substantialinitial errors indicate that the polar axis requires fine adjustments. Improved pointing accuracy is expected to follow subsequent polar alignment adjustments.

The right panel of Figure \ref{fig:pointing_compare} shows the pointing errors after applying the established pointing model. The pointing accuracy is significantly improved with a reduced RMS error of 57.01 arcsec, a median error of 36.91 arcsec and a maximum error of 198.34 arcsec. These results are acceptable given the system’s wide FoV, which is $\sim$18\,deg$^2$. Moreover, the comprehensive use of pointing error data from all selected positions is expected to significantly suppress the largest errors. The large maximum error was likely exacerbated by missing observations in regions affected by cloud coverage, and incorporating a more complete dataset is expected to enhance the robustness of the pointing model.

\subsection{Camera Performance} \label{subsec:camera_performance}

We performed stability tests of the cooling temperature and bias level for two CMOS detectors. During the test, Camera~1 required approximately 1 hour to cool down to $-20\,^{\circ}\mathrm{C}$, whereas Camera~2 required about 3.5 hours to cool down to $-15\,^{\circ}\mathrm{C}$. Consequently, we set the target lock-in temperature as $-20\,^{\circ}\mathrm{C}$ for Camera~1 and $-15\,^{\circ}\mathrm{C}$ for Camera~2. The detector SDK reports temperature values with an integer resolution of $1^{\circ}\mathrm{C}$. Throughout the test, Camera~1 consistently maintained at a temperature of $-20\,^{\circ}\mathrm{C}$, while Camera~2 stabilized at a temperature of $-14^{\circ}\mathrm{C}$, 
demonstrating their excellent thermal stabilities. 

At these stabilized temperatures, the mean bias levels and standard deviations (stdev) were measured as follows: for Camera~1, a mean value of 101.0~ADU is obtained, with a stdev of 1.9~ADU; for Camera~2, the mean value and the stdev are 100.0~ADU and 1.9~ADU, respectively. The two detectors showed highly consistent bias characteristics. 

\begin{figure*}[htbp]
  \centering
  \includegraphics[width=\linewidth]{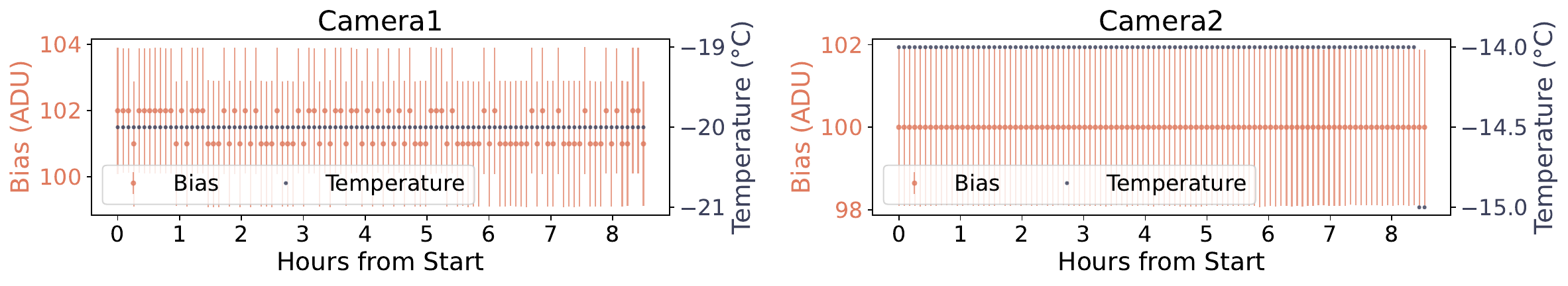}  
\caption{Temporal variations of detector bias level (red points with error bars) and temperature (black line) during a $\sim$9-hour stability test for Camera~1 (left) and Camera~2 (right). 
  The test was conducted at the observatory site, where ambient conditions differ from those in the laboratory. 
  Due to a cooling limitation, the detector temperature of Camera~2 reached only about $-15^{\circ}\mathrm{C}$, while Camera~1 maintained $-20^{\circ}\mathrm{C}$. 
  Bias values represent the median signal within the central $500\times500$ pixel region, with error bars indicating standard deviations.}

  \label{fig:bias_temp}
\end{figure*}

Figure~\ref{fig:bias_temp} shows the variations in detector temperature and bias recorded over a period of about 9 hours. The red data points with error bars, sampled every 5 minutes, represent the median bias and stdev calculated from the central $500\times500$ pixel region of the detectors. The blue data points indicate the corresponding temperature readings, also sampled at a 5-minute interval. The results reveal stable detector temperatures ($-20^{\circ}\mathrm{C}$ for Camera~1 and $\sim$$-15^{\circ}\mathrm{C}$ for Camera~2) and highly consistent bias levels as well. 

We measured the readout noise and dark current of two cameras in two modes. The detector was operated at $-25^\circ\mathrm{C}$ in complete darkness using a light-tight cover. Bias frames at the camera's shortest exposure (0~ms) and dark frames with 500~s integration were acquired. The choice of $-25^\circ\mathrm{C}$ was motivated by stable laboratory environment during the test,
in contrast to the operational settings of $-20^\circ\mathrm{C}$ and $-15^\circ\mathrm{C}$ adopted on site due to different camera cooling times. When determining the readout noise, 50 bias frames were randomized and paired into 25 difference images, with each being clipped to remove outliers. After computing their stdev $\sigma_i$ and taking the median $\sigma_{\rm med}$, the readout-noise was calculated as
\[
  \sigma_{\rm read}
  = \frac{G\,\sigma_{\rm med}}{\sqrt{2}},
\]
where $G$ is the system gain in electrons per analog-to-digital unit (e$^-$/ADU). For dark-current measurement, the average bias and dark frames were subtracted to form a dark field. Within a uniform region after removing hot pixels with $5\sigma$-clipping, the mean pixel value $\bar N_{\rm dark-bias}$ (in ADU) yielded
\[
  \bar I_{\rm dark}
  = \frac{\bar N_{\rm dark-bias}\,G}{t_{\rm exp}},
  \quad t_{\rm exp}=500\,\mathrm{s}.
\]

The test results are presented in Table \ref{tbl:camera-test}.

\begin{deluxetable}{llcc}
\tabletypesize{\scriptsize}
\tablewidth{0pt} 
\tablecaption{Test results of two cameras in two readout modes.\label{tbl:camera-test}}
\tablehead{
\colhead{Camera} & 
\colhead{Parameter} & 
\colhead{High-speed} & 
\colhead{Low-noise}
}
\startdata
1 & Readout noise (e$^-$) & 1.6 & 8.3 \\
  & Dark current (e$^-\,\mathrm{pix}^{-1}\,\mathrm{s}^{-1}$) & 0.00 & 0.066 \\
  & Gain ($e^-$/ADU) & 0.93 & 1.70 \\
  & Full well depth ($e^-$) & $1.7\times10^{4}$ & $1.19\times10^{5}$ \\
2 & Readout noise (e$^-$) & 1.5 & 7.3 \\
  & Dark current (e$^-\,\mathrm{pix}^{-1}\,\mathrm{s}^{-1}$) & 0.00 & 0.11 \\
  & Gain ($e^-$/ADU) & 0.93 & 1.73 \\
  & Full well depth ($e^-$) & $1.6\times10^{4}$ & $1.30\times10^{5}$ \\
\enddata
\tablecomments{
The high-speed mode uses a 14-bit ADC, while the low-noise mode employs a 16-bit ADC.
Differences in readout noise reflect distinct analog/digital readout paths.
Small variations in dark current are within measurement uncertainty.
}
\end{deluxetable}

To characterize the detector’s quantum efficiency (QE), the monochromator output was coupled to an integrating sphere and directed onto the detector through the camera’s optical window. At each wavelength, the exposure time was adjusted so that the resulting flat‐field image filled roughly half of the full‐scale range. Measurements were taken with a 20~nm interval across the wavelength range from 250 to 1010 nm, recording one flat‐field image per wavelength. At the same time, a calibrated monitor photodiode, traceable to a standard reference and cross-calibrated against an identical device placed at the telescope focal plane, continuously tracked the sphere output and provided the absolute irradiance. The digital counts from the detectors were converted to electrons using the known system gain and then compared to the number of incident photons as derived from the monitor current and its responsivity. The ratio of generated electrons to incident photons at each wavelength defines the QE curve.

\begin{figure}[htbp]
  \centering
  \includegraphics[width=\linewidth]{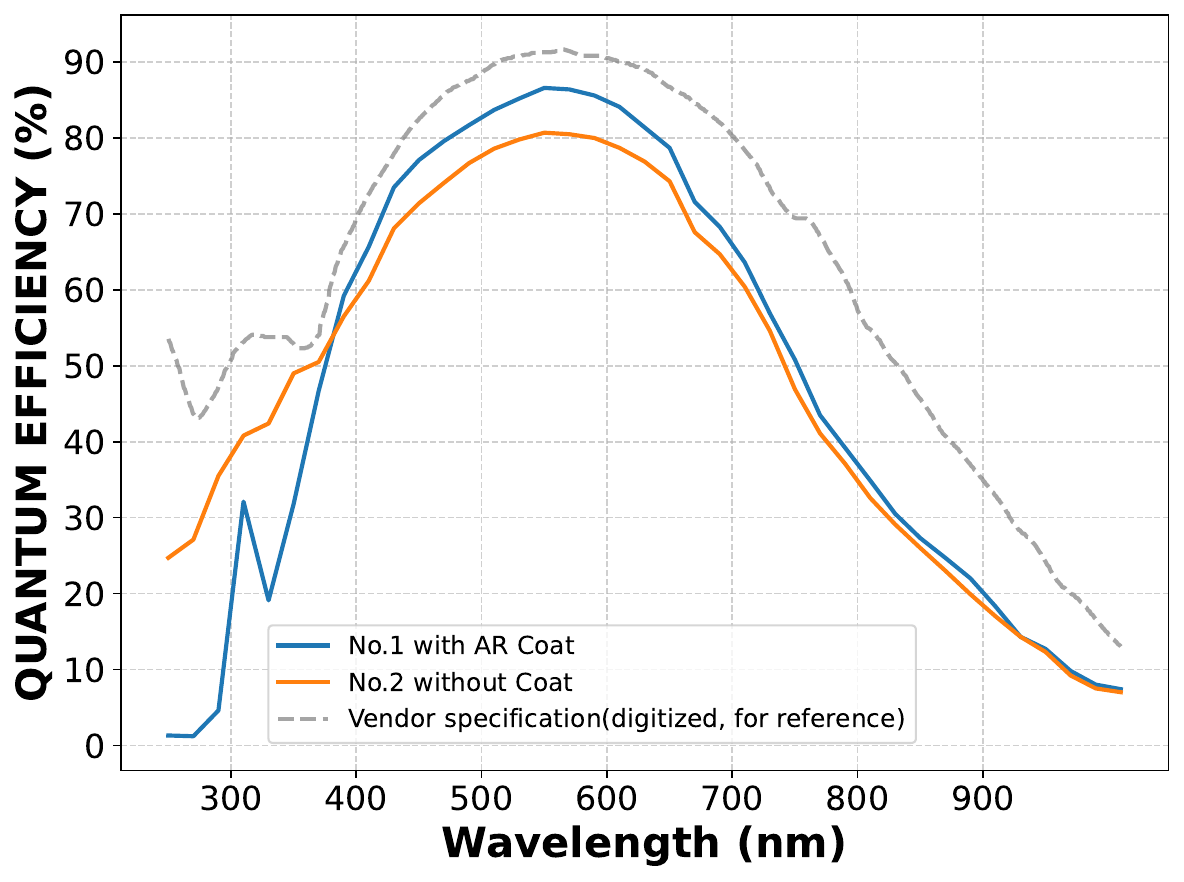}  
\caption{Measured QE curves of the two sCMOS detectors. 
Detector~No.~1 (blue) is equipped with an AR coating, while Detector~No.~2 (orange) is uncoated. The dashed gray curve shows the vendor-specified QE curve of the COSMOS camera \citep{TeledyneCOSMOS}, included for reference. The QE was measured for wavelength ranging from 250 to 1010~nm, with a 20~nm steps using monochromator illumination through an integrating sphere, with absolute irradiance monitored by a calibrated photodiode. 
The AR-coated detector shows enhanced sensitivity across the 400--700~nm range, achieving a peak QE of nearly 90\%, compared to $\sim$78\% for the uncoated one.}

  \label{fig:qe}
\end{figure}

The PI COSMOS camera is available with different anti-reflection (AR) coatings. Detector~No.~1 was fabricated with an AR coating to enhance sensitivity in the 400--700~nm range, while Detector~No.~2 was left uncoated. Their measured QE curves are presented in Figure~\ref{fig:qe}.

\subsection{Limiting Magnitude} \label{subsec:magnitude_limit}

The limiting magnitude depends on multiple factors, including atmospheric conditions, site environment, telescope and camera characteristics, observational strategy, and data reduction procedures. During the commissioning phase, relatively high-quality images were selected and the central $2500\times2500$ pixel region was cropped to reduce optical aberrations and vignetting effects near the edges of the images. Source detection and photometry were performed with \texttt{SExtractor} using the \texttt{MAG\_AUTO} and \texttt{MAGERR\_AUTO} parameters. Finally, the photometry was calibrated against the Pan-STARRS catalog. 
\begin{figure}[htbp]
  \centering
  \includegraphics[width=\linewidth]{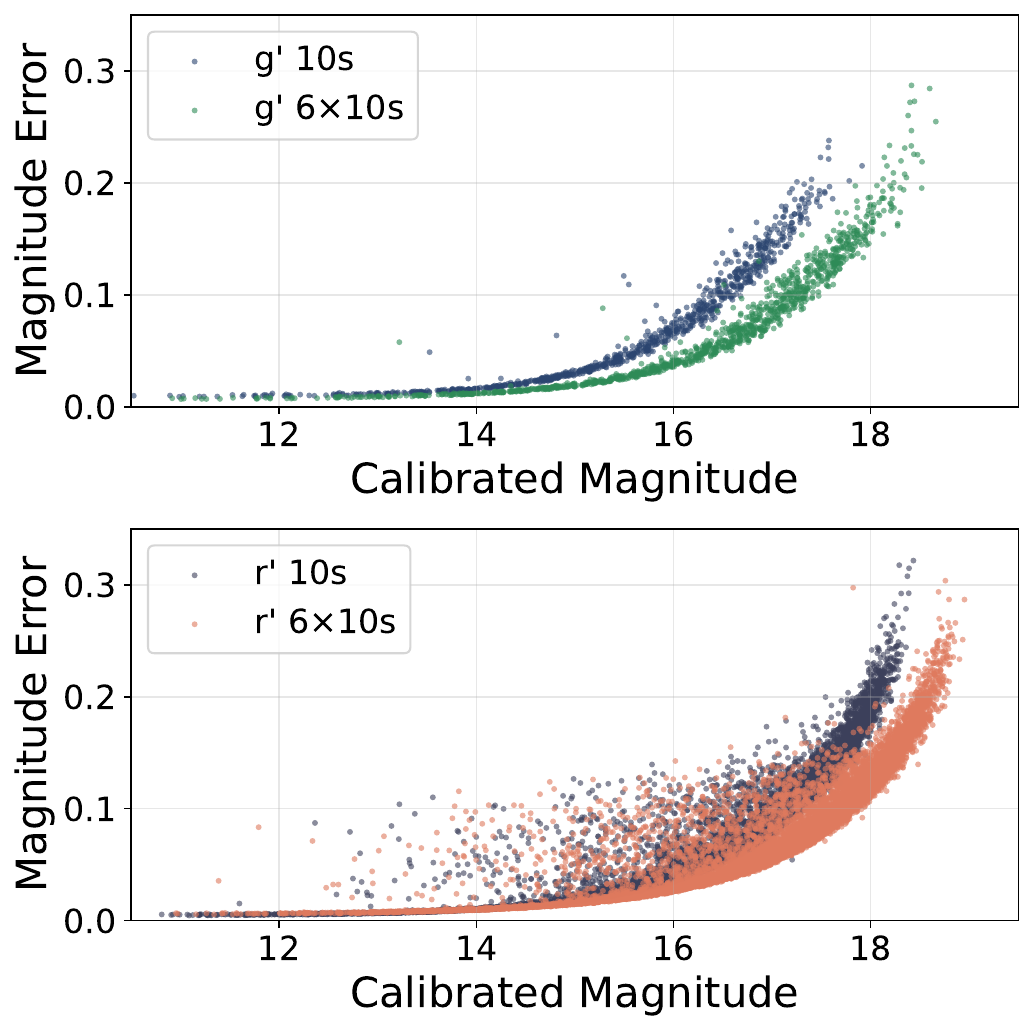}  
\caption{Photometric error as a function of calibrated magnitude in the $g'$ (top) and $r'$ (bottom) bands. 
Results are shown for single 10~s exposures and a stack of 6 consecutive 10~s frames. 
The corresponding 5$\sigma$ limiting magnitudes are 17.5~mag and 18.5~mag in $g'$, and 18.2~mag and 18.7~mag in $r'$, respectively.}

  \label{fig:magnitude_limit}
\end{figure}

Figure \ref{fig:magnitude_limit} shows the distributions of  calibrated photometric magnitudes and errors for the $g'$ and $r'$ bands. In the $g'$ band, the 5$\sigma$ limiting magnitude reaches 17.5~mag for a single 10~s exposure and increases to 18.5~mag when stacking six 10~s exposures. In the $r'$ band, the corresponding limits are 18.2~mag (for a single 10~s) and 18.7~mag (for a stack of six 10~s frames). Commissioning of the R2Pub system is ongoing, and further improvements in detection limit are anticipated as the performance and observing strategies of the system are refined.

\subsection{Initial Scientific Results} \label{subsec:r2pub_results}
\begin{figure}[htbp]
  \centering
  \includegraphics[width=1.1\linewidth]{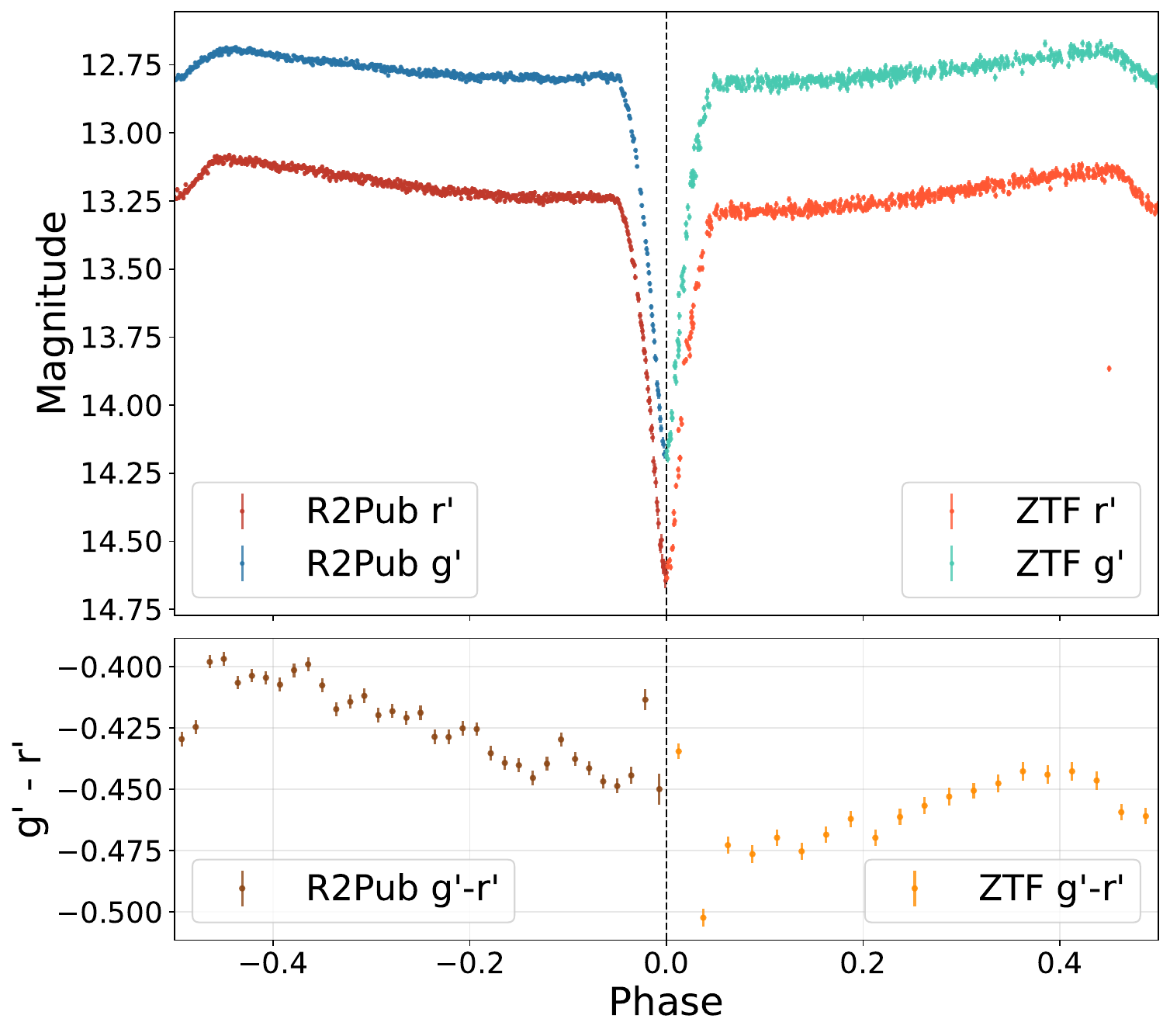}  
\caption{
Phase-folded light curves of the HW~Vir-type eclipsing binary obtained with R2Pub (top left: $g^\prime$, $r^\prime$) and ZTF (top right: $g^\prime$, $r^\prime$). 
The lower panels show the corresponding $g^\prime-r^\prime$ color variations as a function of orbital phase. 
The R2Pub dataset exhibits smoother curves and smaller scatter ($\sim$0.01 mag) due to higher photometric precision and denser sampling, whereas the ZTF data show larger dispersion ($\sim$0.02–0.03 mag), particularly around eclipse ingress and egress. 
A systematic offset of $\sim$0.03 mag is visible between the two datasets, likely caused by filter transmission curves, photometric calibration procedures, or atmospheric extinction corrections. 
}
  \label{fig:r2pub_ztf}
\end{figure}

During the commissioning phase, R2Pub conducted 10s-cadence observations over selected sky regions described in Section~\ref{subsec:skycoverage}, with the goal of capturing short-period variable stars and transient sources. The two tubes of R2Pub system were equipped with $g'$ and $r'$ filters, respectively, thus the observations were conducted simultaneously in both $g'$ and $r'$ bands and typically lasted 2 to 4 hours. To facilitate a direct comparison of the photometry between R2Pub and other sky surveys such as TMTS, we also observed several sky regions that had been previously covered by TMTS and LAMOST. 
Various types of variable stars, including eclipsing binaries and Delta Scuti-type pulsators, are detected throughout the observations. The high time-resolution sampling of R2Pub enables detailed characterization of short-period variability, particularly when combining with radial velocity information provided by the LAMOST. 
An HW~Vir-type eclipsing binary was identified during the  commissioning observations. Figure~\ref{fig:r2pub_ztf} shows the R2Pub light curves of this source in comparison with data from the ZTF survey. To evaluate the photometric precision and color variation, we constructed phase-folded light curves and $g^\prime-r^\prime$ color curves using an adaptive binning approach that accounts for different cadences of the two datasets. 

The R2Pub observations provide substantially higher photometric precision and higher temporal sampling, producing smoother light curves and more stable color measurements with a scatter of $\sim$0.01 mag. In contrast, the ZTF data, while offering excellent long-term temporal coverage, exhibit larger photometric scatter ($\sim$0.02–0.03 mag) in the phase-folded light curves, particularly near eclipse ingress and egress where the colors change rapidly.

A systematic offset of about 0.03 mag in the $g^\prime-r^\prime$ color is observed between the two datasets, with ZTF consistently appearing bluer. This offset may arise from differences in filter transmission curves, photometric calibration procedures, or atmospheric extinction corrections, and warrants further investigation for precision applications.

These results highlight the complementary strengths of the two systems: the simultaneous dual-band imaging of R2Pub enables high-precision color measurements in short timescales, while ZTF provides extended temporal coverage for long-term monitoring and detection of secular variations.

\begin{figure}[htbp]
  \centering
  \includegraphics[width=\linewidth]{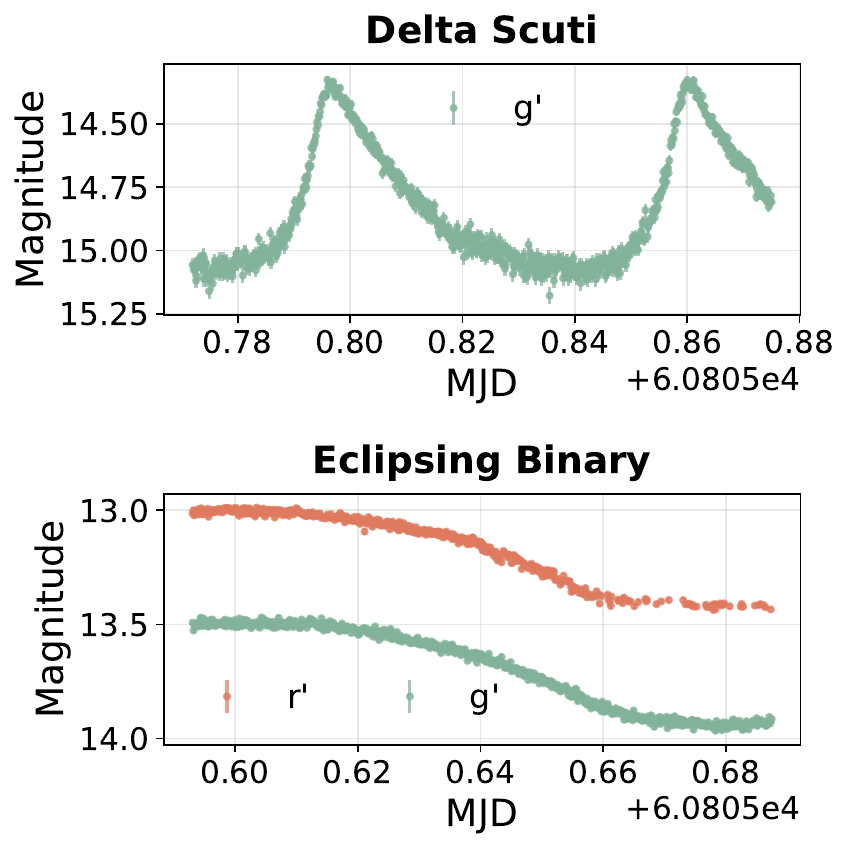}  
\caption{
Typical light curves obtained with the R2Pub during its commissioning phase. 
\textbf{Top:} A $\delta$~Scuti variable observed in the $g^\prime$ band, showing rapid pulsations with a timescale of tens of minutes. 
\textbf{Bottom:} An eclipsing binary observed simultaneously in the $g^\prime$ and $r^\prime$ bands, displaying a primary eclipse with a depth of $\sim$0.5 mag. 
The high-cadence ($10$~s exposures) and continuous coverage of several hours enable the R2Pub system to resolve short-period variability with high precision. 
}
  \label{fig:show_lc}
\end{figure}

Figure~\ref{fig:show_lc} presents representative light curves of a $\delta$~Scuti star and an eclipsing binary taken by R2Pub. The $\delta$~Scuti example shows multi-cycle pulsations on timescales of tens of minutes, while the eclipsing binary exhibits a primary eclipse with a depth of $\sim$0.5~mag. The high temporal resolution of R2Pub observations provides dense phase coverage, allowing us to resolve rapid brightness variations and identify short-period systems. These results demonstrate the system’s capability for time-domain surveys and highlight its potential for discovering and characterizing variable stars and compact binaries.

\section{Summary} \label{sec:summary}
In this paper, we have presented the R2Pub telescope system and its initial scientific performance. The system is designed to conduct a wide-field supernova and short-period variable survey in the $g'$ and $r'$ bands, with an FoV of 18 square degrees. It consists of two telescopes, each equipped with a PI COSMOS camera, and is located at the Daocheng Observatory in Sichuan Province with an elevation of about 4,700\,m.

During the commissioning phase of R2Pub, the system has conducted a pilot survey targeting the TMTS and LAMOST fields. These observations have demonstrated the system’s ability to obtain stable, high-precision photometry across its dual-band configuration, validating its readiness for scientific operations.

The outstanding feature of the R2Pub system is its capability for truly simultaneous observations in two photometric bands. Multi-band synchronized observations are essential in time-domain astronomy, particularly for capturing the early-time color/temperature evolution of transient events. %

To address this requirement, the Beijing Planetarium R2Pub system was developed to provide truly simultaneous dual-band imaging with low-noise, high-speed sCMOS detectors. This capability enables real-time monitoring of rapidly evolving transients and delivers essential color and temperature information for accurate classifications and physical interpretations. 
Additionally, the R2Pub system offers flexible observational modes, including single-band surveys covering $\sim$36\,deg$^{2}$. Enabling both high-cadence monitoring of rapid events and wide-area coverage for transient discovery, significantly enhancing its role in time-domain astronomy.

In conclusion, time-domain astronomy has achieved rapid progress in recent decades. In particular, systems like R2Pub, combining wide-field coverage with simultaneous multi-band capability, will be essential in probing dynamic universe and addressing physical mechanisms of transients such as supernovae, kilonovae, variable stars, AGNs and TDE etc..

\section*{Acknowledgments}
This work was partly supported by the National Natural Science Foundation of China (NSFC, grants 12288102, 12303035, 12203007, 12203006, and 12033003), the Beijing Natural Science Foundation (1242016), and several programs of the Beijing Academy of Science and Technology, including the Young Scholar Program (25CE-YS-02, 24CE-YS-08), the Mengya Program (BGS202203), the Science Program (BS202002), and the Innovation Project (24CD013). This research has also made use of the SIMBAD database, operated at CDS, Strasbourg, France \citep{Wenger2000}.

\software{
Astropy \citep{Astropy2013, Astropy2018},
CCDProc, CuPy,
SExtractor \citep{Bertin1996},
SCAMP \citep{Bertin2006},
SWarp \citep{Bertin2002},
HOTPANTS, Statsmodels, Django, gRPC.
}

\bibliography{PASPsample631}{}

@ARTICLE{2013ApJ...767..143B,
       author = {{Bersten}, Melina C. and {Tanaka}, Masaomi and {Tominaga}, Nozomu and {Benvenuto}, Omar G. and {Nomoto}, Ken'ichi},
        title = "{Early Ultraviolet/Optical Emission of The Type Ib SN 2008D}",
      journal = {\apj},
         year = 2013,
        month = apr,
       volume = {767},
       number = {2},
          eid = {143},
        pages = {143},
          doi = {10.1088/0004-637X/767/2/143},
archivePrefix = {arXiv},
       eprint = {1303.0638},
 primaryClass = {astro-ph.SR},
       adsurl = {https://ui.adsabs.harvard.edu/abs/2013ApJ...767..143B}
}

@INBOOK{2017hsn..book..967W,
	author = {{Waxman}, Eli and {Katz}, Boaz},
	title = "{Shock Breakout Theory}",
	booktitle = {Handbook of Supernovae},
	year = 2017,
	editor = {{Alsabti}, Athem W. and {Murdin}, Paul},
	pages = {967},
	doi = {10.1007/978-3-319-21846-5_33},
	adsurl = {https://ui.adsabs.harvard.edu/abs/2017hsn..book..967W}
}

@ARTICLE{2020PASP..132l5001Z,
       author = {{Zhang}, Ji-Cheng and {Wang}, Xiao-Feng and {Mo}, Jun and {Xi}, Gao-Bo and {Lin}, Jie and {Jiang}, Xiao-Jun and {Zhang}, Xiao-Ming and {Li}, Wen-Xiong and {Yan}, Sheng-Yu and {Chen}, Zhi-Hao and {Hu}, Lei and {Li}, Xue and {Lin}, Wei-Li and {Lin}, Han and {Miao}, Cheng and {Rui}, Li-Ming and {Sai}, Han-Na and {Xiang}, Dan-Feng and {Zhang}, Xing-Han},
        title = "{The Tsinghua University-Ma Huateng Telescopes for Survey: Overview and Performance of the System}",
      journal = {\pasp},
         year = 2020,
        month = dec,
       volume = {132},
       number = {1018},
          eid = {125001},
        pages = {125001},
          doi = {10.1088/1538-3873/abbea2},
archivePrefix = {arXiv},
       eprint = {2012.11456},
 primaryClass = {astro-ph.IM},
       adsurl = {https://ui.adsabs.harvard.edu/abs/2020PASP..132l5001Z}
}

@ARTICLE{2012RAA....12.1197C,
       author = {{Cui}, Xiang-Qun and {Zhao}, Yong-Heng and {Chu}, Yao-Quan and {Li}, Guo-Ping and {Li}, Qi and {Zhang}, Li-Ping and {Su}, Hong-Jun and {Yao}, Zheng-Qiu and {Wang}, Ya-Nan and {Xing}, Xiao-Zheng and {Li}, Xin-Nan and {Zhu}, Yong-Tian and {Wang}, Gang and {Gu}, Bo-Zhong and {Luo}, A. -Li and {Xu}, Xin-Qi and {Zhang}, Zhen-Chao and {Liu}, Gen-Rong and {Zhang}, Hao-Tong and {Yang}, De-Hua and {Cao}, Shu-Yun and {Chen}, Hai-Yuan and {Chen}, Jian-Jun and {Chen}, Kun-Xin and {Chen}, Ying and {Chu}, Jia-Ru and {Feng}, Lei and {Gong}, Xue-Fei and {Hou}, Yong-Hui and {Hu}, Hong-Zhuan and {Hu}, Ning-Sheng and {Hu}, Zhong-Wen and {Jia}, Lei and {Jiang}, Fang-Hua and {Jiang}, Xiang and {Jiang}, Zi-Bo and {Jin}, Ge and {Li}, Ai-Hua and {Li}, Yan and {Li}, Ye-Ping and {Liu}, Guan-Qun and {Liu}, Zhi-Gang and {Lu}, Wen-Zhi and {Mao}, Yin-Dun and {Men}, Li and {Qi}, Yong-Jun and {Qi}, Zhao-Xiang and {Shi}, Huo-Ming and {Tang}, Zheng-Hong and {Tao}, Qing-Sheng and {Wang}, Da-Qi and {Wang}, Dan and {Wang}, Guo-Min and {Wang}, Hai and {Wang}, Jia-Ning and {Wang}, Jian and {Wang}, Jian-Ling and {Wang}, Jian-Ping and {Wang}, Lei and {Wang}, Shu-Qing and {Wang}, You and {Wang}, Yue-Fei and {Xu}, Ling-Zhe and {Xu}, Yan and {Yang}, Shi-Hai and {Yu}, Yong and {Yuan}, Hui and {Yuan}, Xiang-Yan and {Zhai}, Chao and {Zhang}, Jing and {Zhang}, Yan-Xia and {Zhang}, Yong and {Zhao}, Ming and {Zhou}, Fang and {Zhou}, Guo-Hua and {Zhu}, Jie and {Zou}, Si-Cheng},
        title = "{The Large Sky Area Multi-Object Fiber Spectroscopic Telescope (LAMOST)}",
      journal = {Research in Astronomy and Astrophysics},
         year = 2012,
        month = sep,
       volume = {12},
       number = {9},
        pages = {1197-1242},
          doi = {10.1088/1674-4527/12/9/003},
       adsurl = {https://ui.adsabs.harvard.edu/abs/2012RAA....12.1197C}
}

@article{Han_2025,
doi = {10.1088/1674-4527/adc791},
url = {https://dx.doi.org/10.1088/1674-4527/adc791},
year = {2025},
month = {may},
publisher = {National Astromonical Observatories, CAS and IOP Publishing},
volume = {25},
number = {4},
pages = {044009},
author = {Han, Henggeng and Huang, Yang and Wang, Beichuan and Sun, Yongkang and Wang, Cunshi and Li, Zhirui and Jin, Junjie and Sun, Ningchen and Xiao, Kai and He, Min and Gu, Hongrui and Niu, Zexi and Wu, Hong and Liu, Jifeng},
title = {The Mini-SiTian Array: White Paper},
journal = {Research in Astronomy and Astrophysics}
}

@ARTICLE{Liu2021AABC,
  author  = {{Liu}, Jifeng and {Soria}, Roberto and {Wu}, Xue-Feng and {Wu}, Hong and {Shang}, Zhaohui},
  title   = "{The {SiTian} Project}",
  journal = {Anais da Academia Brasileira de Ci\^encias},
  year    = 2021,
  volume  = {93},
  number  = {suppl 1},
  pages   = {e20200628},
  doi     = {10.1590/0001-3765202120200628}
}

@ARTICLE{2019PASP..131a8002B,
       author = {{Bellm}, Eric C. and {Kulkarni}, Shrinivas R. and {Graham}, Matthew J. and {Dekany}, Richard and {Smith}, Roger M. and {Riddle}, Reed and {Masci}, Frank J. and {Helou}, George and {Prince}, Thomas A. and {Adams}, Scott M. and {Barbarino}, C. and {Barlow}, Tom and {Bauer}, James and {Beck}, Ron and {Belicki}, Justin and {Biswas}, Rahul and {Blagorodnova}, Nadejda and {Bodewits}, Dennis and {Bolin}, Bryce and {Brinnel}, Valery and {Brooke}, Tim and {Bue}, Brian and {Bulla}, Mattia and {Burruss}, Rick and {Cenko}, S. Bradley and {Chang}, Chan-Kao and {Connolly}, Andrew and {Coughlin}, Michael and {Cromer}, John and {Cunningham}, Virginia and {De}, Kishalay and {Delacroix}, Alex and {Desai}, Vandana and {Duev}, Dmitry A. and {Eadie}, Gwendolyn and {Farnham}, Tony L. and {Feeney}, Michael and {Feindt}, Ulrich and {Flynn}, David and {Franckowiak}, Anna and {Frederick}, S. and {Fremling}, C. and {Gal-Yam}, Avishay and {Gezari}, Suvi and {Giomi}, Matteo and {Goldstein}, Daniel A. and {Golkhou}, V. Zach and {Goobar}, Ariel and {Groom}, Steven and {Hacopians}, Eugean and {Hale}, David and {Henning}, John and {Ho}, Anna Y.~Q. and {Hover}, David and {Howell}, Justin and {Hung}, Tiara and {Huppenkothen}, Daniela and {Imel}, David and {Ip}, Wing-Huen and {Ivezi{\'c}}, {\v{Z}}eljko and {Jackson}, Edward and {Jones}, Lynne and {Juric}, Mario and {Kasliwal}, Mansi M. and {Kaspi}, S. and {Kaye}, Stephen and {Kelley}, Michael S.~P. and {Kowalski}, Marek and {Kramer}, Emily and {Kupfer}, Thomas and {Landry}, Walter and {Laher}, Russ R. and {Lee}, Chien-De and {Lin}, Hsing Wen and {Lin}, Zhong-Yi and {Lunnan}, Ragnhild and {Giomi}, Matteo and {Mahabal}, Ashish and {Mao}, Peter and {Miller}, Adam A. and {Monkewitz}, Serge and {Murphy}, Patrick and {Ngeow}, Chow-Choong and {Nordin}, Jakob and {Nugent}, Peter and {Ofek}, Eran and {Patterson}, Maria T. and {Penprase}, Bryan and {Porter}, Michael and {Rauch}, Ludwig and {Rebbapragada}, Umaa and {Reiley}, Dan and {Rigault}, Mickael and {Rodriguez}, Hector and {van Roestel}, Jan and {Rusholme}, Ben and {van Santen}, Jakob and {Schulze}, S. and {Shupe}, David L. and {Singer}, Leo P. and {Soumagnac}, Maayane T. and {Stein}, Robert and {Surace}, Jason and {Sollerman}, Jesper and {Szkody}, Paula and {Taddia}, F. and {Terek}, Scott and {Van Sistine}, Angela and {van Velzen}, Sjoert and {Vestrand}, W. Thomas and {Walters}, Richard and {Ward}, Charlotte and {Ye}, Quan-Zhi and {Yu}, Po-Chieh and {Yan}, Lin and {Zolkower}, Jeffry},
        title = "{The Zwicky Transient Facility: System Overview, Performance, and First Results}",
      journal = {\pasp},
         year = 2019,
        month = jan,
       volume = {131},
       number = {995},
        pages = {018002},
          doi = {10.1088/1538-3873/aaecbe},
archivePrefix = {arXiv},
       eprint = {1902.01932},
 primaryClass = {astro-ph.IM},
       adsurl = {https://ui.adsabs.harvard.edu/abs/2019PASP..131a8002B}
}

@ARTICLE{2011PASP..123...58T,
       author = {{Tonry}, John L.},
        title = "{An Early Warning System for Asteroid Impact}",
      journal = {\pasp},
         year = 2011,
        month = jan,
       volume = {123},
       number = {899},
        pages = {58},
          doi = {10.1086/657997},
archivePrefix = {arXiv},
       eprint = {1011.1028},
 primaryClass = {astro-ph.IM},
       adsurl = {https://ui.adsabs.harvard.edu/abs/2011PASP..123...58T}
}

@ARTICLE{2018PASP..130f4505T,
       author = {{Tonry}, J.~L. and {Denneau}, L. and {Heinze}, A.~N. and {Stalder}, B. and {Smith}, K.~W. and {Smartt}, S.~J. and {Stubbs}, C.~W. and {Weiland}, H.~J. and {Rest}, A.},
        title = "{ATLAS: A High-cadence All-sky Survey System}",
      journal = {\pasp},
         year = 2018,
        month = jun,
       volume = {130},
       number = {988},
        pages = {064505},
          doi = {10.1088/1538-3873/aabadf},
archivePrefix = {arXiv},
       eprint = {1802.00879},
 primaryClass = {astro-ph.IM},
       adsurl = {https://ui.adsabs.harvard.edu/abs/2018PASP..130f4505T}
}

@INPROCEEDINGS{2018SPIE10704E..0CD,
       author = {{Dyer}, Martin J. and {Dhillon}, Vik S. and {Littlefair}, Stuart and {Steeghs}, Danny and {Ulaczyk}, Krzysztof and {Chote}, Paul and {Galloway}, Duncan and {Rol}, Evert},
        title = "{A telescope control and scheduling system for the Gravitational-wave Optical Transient Observer (GOTO)}",
    booktitle = {Observatory Operations: Strategies, Processes, and Systems VII},
         year = 2018,
       series = {Society of Photo-Optical Instrumentation Engineers (SPIE) Conference Series},
       volume = {10704},
        month = jul,
          eid = {107040C},
        pages = {107040C},
          doi = {10.1117/12.2311865},
archivePrefix = {arXiv},
       eprint = {1807.01614},
 primaryClass = {astro-ph.IM},
       adsurl = {https://ui.adsabs.harvard.edu/abs/2018SPIE10704E..0CD}
}

@ARTICLE{2009arXiv0912.0201L,
       author = {{LSST Science Collaboration} and {Abell}, Paul A. and {Allison}, Julius and {Anderson}, Scott F. and {Andrew}, John R. and {Angel}, J. Roger P. and {Armus}, Lee and {Arnett}, David and {Asztalos}, S.~J. and {Axelrod}, Tim S. and {Bailey}, Stephen and {Ballantyne}, D.~R. and {Bankert}, Justin R. and {Barkhouse}, Wayne A. and {Barr}, Jeffrey D. and {Barrientos}, L. Felipe and {Barth}, Aaron J. and {Bartlett}, James G. and {Becker}, Andrew C. and {Becla}, Jacek and {Beers}, Timothy C. and {Bernstein}, Joseph P. and {Biswas}, Rahul and {Blanton}, Michael R. and {Bloom}, Joshua S. and {Bochanski}, John J. and {Boeshaar}, Pat and {Borne}, Kirk D. and {Bradac}, Marusa and {Brandt}, W.~N. and {Bridge}, Carrie R. and {Brown}, Michael E. and {Brunner}, Robert J. and {Bullock}, James S. and {Burgasser}, Adam J. and {Burge}, James H. and {Burke}, David L. and {Cargile}, Phillip A. and {Chandrasekharan}, Srinivasan and {Chartas}, George and {Chesley}, Steven R. and {Chu}, You-Hua and {Cinabro}, David and {Claire}, Mark W. and {Claver}, Charles F. and {Clowe}, Douglas and {Connolly}, A.~J. and {Cook}, Kem H. and {Cooke}, Jeff and {Cooray}, Asantha and {Covey}, Kevin R. and {Culliton}, Christopher S. and {de Jong}, Roelof and {de Vries}, Willem H. and {Debattista}, Victor P. and {Delgado}, Francisco and {Dell'Antonio}, Ian P. and {Dhital}, Saurav and {Di Stefano}, Rosanne and {Dickinson}, Mark and {Dilday}, Benjamin and {Djorgovski}, S.~G. and {Dobler}, Gregory and {Donalek}, Ciro and {Dubois-Felsmann}, Gregory and {Durech}, Josef and {Eliasdottir}, Ardis and {Eracleous}, Michael and {Eyer}, Laurent and {Falco}, Emilio E. and {Fan}, Xiaohui and {Fassnacht}, Christopher D. and {Ferguson}, Harry C. and {Fernandez}, Yanga R. and {Fields}, Brian D. and {Finkbeiner}, Douglas and {Figueroa}, Eduardo E. and {Fox}, Derek B. and {Francke}, Harold and {Frank}, James S. and {Frieman}, Josh and {Fromenteau}, Sebastien and {Furqan}, Muhammad and {Galaz}, Gaspar and {Gal-Yam}, A. and {Garnavich}, Peter and {Gawiser}, Eric and {Geary}, John and {Gee}, Perry and {Gibson}, Robert R. and {Gilmore}, Kirk and {Grace}, Emily A. and {Green}, Richard F. and {Gressler}, William J. and {Grillmair}, Carl J. and {Habib}, Salman and {Haggerty}, J.~S. and {Hamuy}, Mario and {Harris}, Alan W. and {Hawley}, Suzanne L. and {Heavens}, Alan F. and {Hebb}, Leslie and {Henry}, Todd J. and {Hileman}, Edward and {Hilton}, Eric J. and {Hoadley}, Keri and {Holberg}, J.~B. and {Holman}, Matt J. and {Howell}, Steve B. and {Infante}, Leopoldo and {Ivezic}, Zeljko and {Jacoby}, Suzanne H. and {Jain}, Bhuvnesh and {R} and {Jedicke} and {Jee}, M. James and {Garrett Jernigan}, J. and {Jha}, Saurabh W. and {Johnston}, Kathryn V. and {Jones}, R. Lynne and {Juric}, Mario and {Kaasalainen}, Mikko and {Styliani} and {Kafka} and {Kahn}, Steven M. and {Kaib}, Nathan A. and {Kalirai}, Jason and {Kantor}, Jeff and {Kasliwal}, Mansi M. and {Keeton}, Charles R. and {Kessler}, Richard and {Knezevic}, Zoran and {Kowalski}, Adam and {Krabbendam}, Victor L. and {Krughoff}, K. Simon and {Kulkarni}, Shrinivas and {Kuhlman}, Stephen and {Lacy}, Mark and {Lepine}, Sebastien and {Liang}, Ming and {Lien}, Amy and {Lira}, Paulina and {Long}, Knox S. and {Lorenz}, Suzanne and {Lotz}, Jennifer M. and {Lupton}, R.~H. and {Lutz}, Julie and {Macri}, Lucas M. and {Mahabal}, Ashish A. and {Mandelbaum}, Rachel and {Marshall}, Phil and {May}, Morgan and {McGehee}, Peregrine M. and {Meadows}, Brian T. and {Meert}, Alan and {Milani}, Andrea and {Miller}, Christopher J. and {Miller}, Michelle and {Mills}, David and {Minniti}, Dante and {Monet}, David and {Mukadam}, Anjum S. and {Nakar}, Ehud and {Neill}, Douglas R. and {Newman}, Jeffrey A. and {Nikolaev}, Sergei and {Nordby}, Martin and {O'Connor}, Paul and {Oguri}, Masamune and {Oliver}, John and {Olivier}, Scot S. and {Olsen}, Julia K. and {Olsen}, Knut and {Olszewski}, Edward W. and {Oluseyi}, Hakeem and {Padilla}, Nelson D. and {Parker}, Alex and {Pepper}, Joshua and {Peterson}, John R. and {Petry}, Catherine and {Pinto}, Philip A. and {Pizagno}, James L. and {Popescu}, Bogdan and {Prsa}, Andrej and {Radcka}, Veljko and {Raddick}, M. Jordan and {Rasmussen}, Andrew and {Rau}, Arne and {Rho}, Jeonghee and {Rhoads}, James E. and {Richards}, Gordon T. and {Ridgway}, Stephen T. and {Robertson}, Brant E. and {Roskar}, Rok and {Saha}, Abhijit and {Sarajedini}, Ata and {Scannapieco}, Evan and {Schalk}, Terry and {Schindler}, Rafe and {Schmidt}, Samuel},
        title = "{LSST Science Book, Version 2.0}",
      journal = {arXiv e-prints},
         year = 2009,
        month = dec,
          eid = {arXiv:0912.0201},
        pages = {arXiv:0912.0201},
          doi = {10.48550/arXiv.0912.0201},
archivePrefix = {arXiv},
       eprint = {0912.0201},
 primaryClass = {astro-ph.IM},
       adsurl = {https://ui.adsabs.harvard.edu/abs/2009arXiv0912.0201L}
}

@ARTICLE{1998ApJ...507L..59L,
       author = {{Li}, Li-Xin and {Paczy{\'n}ski}, Bohdan},
        title = "{Transient Events from Neutron Star Mergers}",
      journal = {\apjl},
         year = 1998,
        month = nov,
       volume = {507},
       number = {1},
        pages = {L59-L62},
          doi = {10.1086/311680},
archivePrefix = {arXiv},
       eprint = {astro-ph/9807272},
 primaryClass = {astro-ph},
       adsurl = {https://ui.adsabs.harvard.edu/abs/1998ApJ...507L..59L}
}

@ARTICLE{2010MNRAS.406.2650M,
       author = {{Metzger}, B.~D. and {Mart{\'\i}nez-Pinedo}, G. and {Darbha}, S. and {Quataert}, E. and {Arcones}, A. and {Kasen}, D. and {Thomas}, R. and {Nugent}, P. and {Panov}, I.~V. and {Zinner}, N.~T.},
        title = "{Electromagnetic counterparts of compact object mergers powered by the radioactive decay of r-process nuclei}",
      journal = {\mnras},
         year = 2010,
        month = aug,
       volume = {406},
       number = {4},
        pages = {2650-2662},
          doi = {10.1111/j.1365-2966.2010.16864.x},
archivePrefix = {arXiv},
       eprint = {1001.5029},
 primaryClass = {astro-ph.HE},
       adsurl = {https://ui.adsabs.harvard.edu/abs/2010MNRAS.406.2650M}
}

@ARTICLE{Nomoto1982a,
       author = {{Nomoto}, K.},
        title = "{Accreting white dwarf models for type I supernovae. I - Presupernova evolution and triggering mechanisms}",
      journal = {\apj},
         year = 1982,
        month = feb,
       volume = {253},
        pages = {798-810},
          doi = {10.1086/159682},
       adsurl = {https://ui.adsabs.harvard.edu/abs/1982ApJ...253..798N}
}

@ARTICLE{Nomoto1982b,
       author = {{Nomoto}, K.},
        title = "{Accreting white dwarf models for type I supernovae. II. Off-center detonation supernovae.}",
      journal = {\apj},
         year = 1982,
        month = jun,
       volume = {257},
        pages = {780-792},
          doi = {10.1086/160031},
       adsurl = {https://ui.adsabs.harvard.edu/abs/1982ApJ...257..780N}
}

@ARTICLE{Woosley2011,
       author = {{Woosley}, S.~E. and {Kasen}, Daniel},
        title = "{Sub-Chandrasekhar Mass Models for Supernovae}",
      journal = {\apj},
         year = 2011,
        month = jun,
       volume = {734},
       number = {1},
          eid = {38},
        pages = {38},
          doi = {10.1088/0004-637X/734/1/38},
archivePrefix = {arXiv},
       eprint = {1010.5292},
 primaryClass = {astro-ph.HE},
       adsurl = {https://ui.adsabs.harvard.edu/abs/2011ApJ...734...38W}
}

@ARTICLE{Kasen2010,
       author = {{Kasen}, Daniel},
        title = "{Seeing the Collision of a Supernova with Its Companion Star}",
      journal = {\apj},
         year = 2010,
        month = jan,
       volume = {708},
       number = {2},
        pages = {1025-1031},
          doi = {10.1088/0004-637X/708/2/1025},
archivePrefix = {arXiv},
       eprint = {0909.0275},
 primaryClass = {astro-ph.HE},
       adsurl = {https://ui.adsabs.harvard.edu/abs/2010ApJ...708.1025K}
}

@ARTICLE{Lim2023,
       author = {{Lim}, Gu and {Im}, Myungshin and {Paek}, Gregory S.~H. and {Yoon}, Sung-Chul and {Choi}, Changsu and {Kim}, Sophia and {Wheeler}, J. Craig and {Thomas}, Benjamin P. and {Vink{\'o}}, Jozsef and {Kim}, Dohyeong and {Seo}, Jinguk and {Kang}, Wonseok and {Kim}, Taewoo and {Sung}, Hyun-Il and {Kim}, Yonggi and {Yoon}, Joh-Na and {Kim}, Haeun and {Kim}, Jeongmook and {Bae}, Hana and {Ehgamberdiev}, Shuhrat and {Burhonov}, Otabek and {Mirzaqulov}, Davron},
        title = "{The Early Light Curve of the Type Ia Supernova 2021hpr in NGC 3147: Progenitor Constraints with the Companion Interaction Model}",
      journal = {\apj},
         year = 2023,
        month = may,
       volume = {949},
       number = {1},
          eid = {33},
        pages = {33},
          doi = {10.3847/1538-4357/acc10c},
archivePrefix = {arXiv},
       eprint = {2303.05051},
 primaryClass = {astro-ph.SR},
       adsurl = {https://ui.adsabs.harvard.edu/abs/2023ApJ...949...33L}
}

@ARTICLE{WFST2023,
       author = {{Wang}, Tinggui and {Liu}, Guilin and {Cai}, Zhenyi and {Geng}, Jinjun and {Fang}, Min and {He}, Haoning and {Jiang}, Ji-an and {Jiang}, Ning and {Kong}, Xu and {Li}, Bin and {Li}, Ye and {Luo}, Wentao and {Pan}, Zhizheng and {Wu}, Xuefeng and {Yang}, Ji and {Yu}, Jiming and {Zheng}, Xianzhong and {Zhu}, Qingfeng and {Cai}, Yi-Fu and {Chen}, Yuanyuan and {Chen}, Zhiwei and {Dai}, Zigao and {Fan}, Lulu and {Fan}, Yizhong and {Fang}, Wenjuan and {He}, Zhicheng and {Hu}, Lei and {Hu}, Maokai and {Jin}, Zhiping and {Jiang}, Zhibo and {Li}, Guoliang and {Li}, Fan and {Li}, Xuzhi and {Liang}, Runduo and {Lin}, Zheyu and {Liu}, Qingzhong and {Liu}, Wenhao and {Liu}, Zhengyan and {Liu}, Wei and {Liu}, Yao and {Lou}, Zheng and {Qu}, Han and {Sheng}, Zhenfeng and {Shi}, Jianchun and {Shu}, Yiping and {Su}, Zhenbo and {Sun}, Tianrui and {Wang}, Hongchi and {Wang}, Huiyuan and {Wang}, Jian and {Wang}, Junxian and {Wei}, Daming and {Wei}, Junjie and {Xue}, Yongquan and {Yan}, Jingzhi and {Yang}, Chao and {Yuan}, Ye and {Yuan}, Yefei and {Zhang}, Hongxin and {Zhang}, Miaomiao and {Zhao}, Haibin and {Zhao}, Wen},
        title = "{Science with the 2.5-meter Wide Field Survey Telescope (WFST)}",
      journal = {Science China Physics, Mechanics, and Astronomy},
         year = 2023,
        month = oct,
       volume = {66},
       number = {10},
          eid = {109512},
        pages = {109512},
          doi = {10.1007/s11433-023-2197-5},
archivePrefix = {arXiv},
       eprint = {2306.07590},
 primaryClass = {astro-ph.IM},
       adsurl = {https://ui.adsabs.harvard.edu/abs/2023SCPMA..6609512W}
}

@ARTICLE{songtengfei2020,
       author = {{Song}, Teng-Fei and {Liu}, Yu and {Wang}, Jing-Xing and {Zhang}, Xue-Fei and {Liu}, Shun-Qing and {Zhao}, Ming-Yu and {Li}, Xiao-Bo and {Cai}, Zhan-Chuan and {Song}, Qi-Wu and {Cao}, Zi-Huang and {Ruan}, Yu},
        title = "{Site testing campaign for the Large Optical/infrared Telescope of China: general introduction of the Daocheng site}",
      journal = {Research in Astronomy and Astrophysics},
         year = 2020,
        month = jun,
       volume = {20},
       number = {6},
          eid = {085},
        pages = {085},
          doi = {10.1088/1674-4527/20/6/85},
       adsurl = {https://ui.adsabs.harvard.edu/abs/2020RAA....20...85S}
}

@article{Brennan_2022,
   title={The Automated Photometry of Transients pipeline (AutoPhOT)},
   volume={667},
   ISSN={1432-0746},
   url={http://dx.doi.org/10.1051/0004-6361/202243067},
   DOI={10.1051/0004-6361/202243067},
   journal={Astronomy \& Astrophysics},
   publisher={EDP Sciences},
   author={Brennan, S. J. and Fraser, M.},
   year={2022},
   month=nov, pages={A62} }

@software{stdpipe,
       author = {{Karpov}, Sergey},
        title = "{STDPipe: Simple Transient Detection Pipeline}",
 howpublished = {Astrophysics Source Code Library, record ascl:2112.006},
         year = 2021,
        month = dec,
          eid = {ascl:2112.006},
       adsurl = {https://ui.adsabs.harvard.edu/abs/2021ascl.soft12006K}
}

@ARTICLE{Kochanek2017,
       author = {{Kochanek}, C.~S. and {Shappee}, B.~J. and {Stanek}, K.~Z. and {Holoien}, T.~W. -S. and {Thompson}, Todd A. and {Prieto}, J.~L. and {Dong}, Subo and {Shields}, J.~V. and {Will}, D. and {Britt}, C. and {Perzanowski}, D. and {Pojma{\'n}ski}, G.},
        title = "{The All-Sky Automated Survey for Supernovae (ASAS-SN) Light Curve Server v1.0}",
      journal = {\pasp},
         year = 2017,
        month = oct,
       volume = {129},
       number = {980},
        pages = {104502},
          doi = {10.1088/1538-3873/aa80d9},
archivePrefix = {arXiv},
       eprint = {1706.07060},
 primaryClass = {astro-ph.SR},
       adsurl = {https://ui.adsabs.harvard.edu/abs/2017PASP..129j4502K}
}

@INPROCEEDINGS{Shappee2014,
       author = {{Shappee}, Benjamin and {Prieto}, J. and {Stanek}, K.~Z. and {Kochanek}, C.~S. and {Holoien}, T. and {Jencson}, J. and {Basu}, U. and {Beacom}, J.~F. and {Szczygiel}, D. and {Pojmanski}, G. and {Brimacombe}, J. and {Dubberley}, M. and {Elphick}, M. and {Foale}, S. and {Hawkins}, E. and {Mullins}, D. and {Rosing}, W. and {Ross}, R. and {Walker}, Z.},
        title = "{All Sky Automated Survey for SuperNovae (ASAS-SN or ``Assassin'')}",
    booktitle = {American Astronomical Society Meeting Abstracts \#223},
         year = 2014,
       series = {American Astronomical Society Meeting Abstracts},
       volume = {223},
        month = jan,
          eid = {236.03},
        pages = {236.03},
       adsurl = {https://ui.adsabs.harvard.edu/abs/2014AAS...22323603S}
}

@ARTICLE{nature2024ixf,
       author = {{Li}, Gaici and {Hu}, Maokai and {Li}, Wenxiong and {Yang}, Yi and {Wang}, Xiaofeng and {Yan}, Shengyu and {Hu}, Lei and {Zhang}, Jujia and {Mao}, Yiming and {Riise}, Henrik and {Gao}, Xing and {Sun}, Tianrui and {Liu}, Jialian and {Xiong}, Dingrong and {Wang}, Lifan and {Mo}, Jun and {Iskandar}, Abdusamatjan and {Xi}, Gaobo and {Xiang}, Danfeng and {Wang}, Lingzhi and {Sun}, Guoyou and {Zhang}, Keming and {Chen}, Jian and {Lin}, Weili and {Guo}, Fangzhou and {Liu}, Qichun and {Cai}, Guangyao and {Zhou}, Wenjie and {Zhao}, Jingyuan and {Chen}, Jin and {Zheng}, Xin and {Li}, Keying and {Zhang}, Mi and {Xu}, Shijun and {Lyu}, Xiaodong and {Castro-Tirado}, Alberto J. and {Chufarin}, Vasilii and {Potapov}, Nikolay and {Ionov}, Ivan and {Korotkiy}, Stanislav and {Nazarov}, Sergey and {Sokolovsky}, Kirill and {Hamann}, Norman and {Herman}, Eliot},
        title = "{A shock flash breaking out of a dusty red supergiant}",
      journal = {\nat},
         year = 2024,
        month = mar,
       volume = {627},
       number = {8005},
        pages = {754-758},
          doi = {10.1038/s41586-023-06843-6},
archivePrefix = {arXiv},
       eprint = {2311.14409},
 primaryClass = {astro-ph.HE},
       adsurl = {https://ui.adsabs.harvard.edu/abs/2024Natur.627..754L}
}

@ARTICLE{Wenger2000,
       author = {{Wenger}, M. and {Ochsenbein}, F. and {Egret}, D. and {Dubois}, P. and {Bonnarel}, F. and {Borde}, S. and {Genova}, F. and {Jasniewicz}, G. and {Lalo{\"e}}, S. and {Lesteven}, S. and {Monier}, R.},
        title = "{The SIMBAD astronomical database. The CDS reference database for astronomical objects}",
      journal = {\aaps},
         year = 2000,
        month = apr,
       volume = {143},
        pages = {9-22},
          doi = {10.1051/aas:2000332},
archivePrefix = {arXiv},
       eprint = {astro-ph/0002110},
 primaryClass = {astro-ph},
       adsurl = {https://ui.adsabs.harvard.edu/abs/2000A\&AS..143....9W}
}

@INPROCEEDINGS{liuyu2018,
       author = {{Liu}, Yu and {Li}, Xiaobo and {Zhang}, Xuefei and {Song}, Tengfei and {Wang}, Jingxing and {Zhao}, Mingyu and {Xia}, Lidong and {Song}, Qiwu},
        title = "{Operation of the astronomical monitoring stations at Mt. Wumingshan}",
    booktitle = {Observatory Operations: Strategies, Processes, and Systems VII},
         year = 2018,
       series = {Society of Photo-Optical Instrumentation Engineers (SPIE) Conference Series},
       volume = {10704},
        month = jul,
          eid = {1070422},
        pages = {1070422},
          doi = {10.1117/12.2309831},
       adsurl = {https://ui.adsabs.harvard.edu/abs/2018SPIE10704E..22L}
}

@ARTICLE{blackgem2024,
       author = {{Groot}, P.~J. and {Bloemen}, S. and {Vreeswijk}, P.~M. and {van Roestel}, J.~C.~J. and {Jonker}, P.~G. and {Nelemans}, G. and {Klein-Wolt}, M. and {Lepoole}, R. and {Pieterse}, D.~L.~A. and {Rodenhuis}, M. and {Boland}, W. and {Haverkorn}, M. and {Aerts}, C. and {Bakker}, R. and {Balster}, H. and {Bekema}, M. and {Dijkstra}, E. and {Dolron}, P. and {Elswijk}, E. and {van Elteren}, A. and {Engels}, A. and {Fokker}, M. and {de Haan}, M. and {Hahn}, F. and {ter Horst}, R. and {Lesman}, D. and {Kragt}, J. and {Morren}, J. and {Nillissen}, H. and {Pessemier}, W. and {Raskin}, G. and {de Rijke}, A. and {Scheers}, L.~H.~A. and {Schuil}, M. and {Timmer}, S.~T. and {Antunes Amaral}, L. and {Arancibia-Rojas}, E. and {Arcavi}, I. and {Blagorodnova}, N. and {Biswas}, S. and {Breton}, R.~P. and {Dawson}, H. and {Dayal}, P. and {De Wet}, S. and {Duffy}, C. and {Faris}, S. and {Fausnaugh}, M. and {Gal-Yam}, A. and {Geier}, S. and {Horesh}, A. and {Johnston}, C. and {Katusiime}, G. and {Kelley}, C. and {Kosakowski}, A. and {Kupfer}, T. and {Leloudas}, G. and {Levan}, A. and {Modiano}, D. and {Mogawana}, O. and {Munday}, J. and {Paice}, J. and {Patat}, F. and {Pelisoli}, I. and {Ramsay}, G. and {Ranaivomanana}, P.~T. and {Ruiz-Carmona}, R. and {Schaffenroth}, V. and {Scaringi}, S. and {Stoppa}, F. and {Street}, R. and {Tranin}, H. and {Uzundag}, M. and {Valenti}, S. and {Veresvarska}, M. and {Vuc̆kovi{\'c}}, M. and {Wichern}, H.~C.~I. and {Wijers}, R.~A.~M.~J. and {Wijnands}, R.~A.~D. and {Zimmerman}, E.},
        title = "{The BlackGEM Telescope Array. I. Overview}",
      journal = {\pasp},
         year = 2024,
        month = nov,
       volume = {136},
       number = {11},
          eid = {115003},
        pages = {115003},
          doi = {10.1088/1538-3873/ad8b6a},
archivePrefix = {arXiv},
       eprint = {2405.18923},
 primaryClass = {astro-ph.IM},
       adsurl = {https://ui.adsabs.harvard.edu/abs/2024PASP..136k5003G}
}

@misc{atlas2,
      title={ATLAS-TEIDE: The next generations of ATLAS units for the Teide Observatory}, 
      author={Javier Licandro and John Tonry and Miguel R. Alarcon and Miquel Serra-Ricart and Larry Denneau},
      year={2023},
      eprint={2302.07954},
      archivePrefix={arXiv},
      primaryClass={astro-ph.IM},
      url={https://arxiv.org/abs/2302.07954}, 
}

@article{tess,
  title={Transiting exoplanet survey satellite},
  author={Ricker, George R and Winn, Joshua N and Vanderspek, Roland and Latham, David W and Bakos, G{\'a}sp{\'a}r {\'A} and Bean, Jacob L and Berta-Thompson, Zachory K and Brown, Timothy M and Buchhave, Lars and Butler, Nathaniel R and others},
  journal={Journal of Astronomical Telescopes, Instruments, and Systems},
  volume={1},
  number={1},
  pages={014003--014003},
  year={2015},
  publisher={Society of Photo-Optical Instrumentation Engineers}
}

@ARTICLE{lin2023,
       author = {{Lin}, Jie and {Wang}, Xiaofeng and {Mo}, Jun and {Xi}, Gaobo and {Filippenko}, Alexei V. and {Yan}, Shengyu and {Brink}, Thomas G. and {Yang}, Yi and {Wu}, Chengyuan and {N{\'e}meth}, P{\'e}ter and {Li}, Gaici and {Guo}, Fangzhou and {Guo}, Jincheng and {Cai}, Yongzhi and {Xiong}, Heran and {Zheng}, WeiKang and {Liu}, Qichun and {Zhang}, Jicheng and {Jiang}, Xiaojun and {Chen}, Liyang and {Xia}, Qiqi and {Peng}, Haowei and {Chen}, Zhihao and {Li}, Wenxiong and {Lin}, Weili and {Xiang}, Danfeng and {Ma}, Xiaoran and {Liu}, Jialian},
        title = "{Minute-cadence observations of the LAMOST fields with the TMTS: II. Catalogues of short-period variable stars from the first 2-yr surveys}",
      journal = {\mnras},
         year = 2023,
        month = aug,
       volume = {523},
       number = {2},
        pages = {2172-2192},
          doi = {10.1093/mnras/stad994},
archivePrefix = {arXiv},
       eprint = {2303.18050},
 primaryClass = {astro-ph.SR},
       adsurl = {https://ui.adsabs.harvard.edu/abs/2023MNRAS.523.2172L}
}

@ARTICLE{lin2022,
       author = {{Lin}, Jie and {Wang}, Xiaofeng and {Mo}, Jun and {Xi}, Gaobo and {Zhang}, Jicheng and {Jiang}, Xiaojun and {Shi}, Jianrong and {Zhang}, Xiaobin and {Zhang}, Xiaoming and {Wei}, Zixuan and {Ye}, Limeng and {Wu}, Chengyuan and {Yan}, Shengyu and {Chen}, Zhihao and {Li}, Wenxiong and {Li}, Xue and {Lin}, Weili and {Lin}, Han and {Sai}, Hanna and {Xiang}, Danfeng and {Zhang}, Xinghan},
        title = "{Minute-cadence observations of the LAMOST fields with the TMTS: I. Methodology of detecting short-period variables and results from the first-year survey}",
      journal = {\mnras},
         year = 2022,
        month = jan,
       volume = {509},
       number = {2},
        pages = {2362-2376},
          doi = {10.1093/mnras/stab2812},
archivePrefix = {arXiv},
       eprint = {2109.11155},
 primaryClass = {astro-ph.IM},
       adsurl = {https://ui.adsabs.harvard.edu/abs/2022MNRAS.509.2362L}
}

@article{Astropy2013,
  author = {{Astropy Collaboration}},
  title  = {Astropy: A community Python package for astronomy},
  journal= {A\&A},
  volume = {558},
  pages  = {A33},
  year   = {2013},
  doi    = {10.1051/0004-6361/201322068}
}

@article{Astropy2018,
  author = {{Astropy Collaboration}},
  title  = {The Astropy Project: Building an Open-science Project and Status of the v2.0 Core Package},
  journal= {AJ},
  volume = {156},
  pages  = {123},
  year   = {2018},
  doi    = {10.3847/1538-3881/aabc4f}
}

@article{Bertin1996,
  author = {Bertin, E. and Arnouts, S.},
  title = {SExtractor: Software for source extraction},
  journal = {A\&AS}, volume = {117}, pages = {393}, year = {1996},
  doi = {10.1051/aas:1996164}
}

@article{Bertin2006,
  author = {Bertin, E.},
  title = {Automatic Astrometric and Photometric Calibration with SCAMP},
  journal = {Astronomical Data Analysis Software and Systems XV}, volume = {351}, pages = {112}, year = {2006}
}

@article{Bertin2002,
  author = {Bertin, E. et al.},
  title = {SWarp: Resampling and Co-adding FITS Images Together},
  journal = {Astronomical Data Analysis Software and Systems XI}, volume = {281}, pages = {228}, year = {2002}
}

@ARTICLE{goto2,
       author = {{Makrygianni}, L. and {Mullaney}, J. and {Dhillon}, V. and {Littlefair}, S. and {Ackley}, K. and {Dyer}, M.~J. and {Lyman}, J. and {Ulaczyk}, K. and {Cutter}, R. and {Mong}, Y. -L. and {Steeghs}, D. and {Galloway}, D.~K. and {O'Brien}, P. and {Ramsay}, G. and {Poshyachinda}, S. and {Kotak}, R. and {Nuttall}, L. and {Pall{\'e}}, E. and {Pollacco}, D. and {Thrane}, E. and {Aukkaravittayapun}, S. and {Awiphan}, S. and {Breton}, R.~P. and {Burhanudin}, U. and {Chote}, P. and {Chrimes}, A. and {Daw}, E. and {Duffy}, C. and {Eyles-Ferris}, R. and {Gompertz}, B. and {Heikkil{\"a}}, T. and {Irawati}, P. and {Kennedy}, M. and {Killestein}, T. and {Levan}, A. and {Marsh}, T. and {Mata-Sanchez}, D. and {Mattila}, S. and {Maund}, J. and {McCormac}, J. and {Mkrtichian}, D. and {Rol}, E. and {Sawangwit}, U. and {Stanway}, E. and {Starling}, R. and {Str{\o}m}, P.~A. and {Tooke}, S. and {Wiersema}, K.},
        title = "{Processing GOTO survey data with the Rubin Observatory LSST Science Pipelines II: Forced Photometry and lightcurves}",
      journal = {\pasa},
         year = 2021,
        month = jun,
       volume = {38},
          eid = {e025},
        pages = {e025},
          doi = {10.1017/pasa.2021.19},
archivePrefix = {arXiv},
       eprint = {2105.05128},
 primaryClass = {astro-ph.IM},
       adsurl = {https://ui.adsabs.harvard.edu/abs/2021PASA...38...25M}
}

@ARTICLE{minisitian3,
       author = {{Wang}, Bei-Chuan and {Jin}, Jun-Jie and {Zhang}, Yu and {Wang}, Yanan and {Wang}, Song and {Gu}, Hong-Rui and {He}, Min and {Mu}, Hai-Yang and {Xiao}, Kai and {Li}, Zhi-Rui and {Fan}, Zhou and {Ge}, Liang and {Tian}, Jian-Feng and {Huang}, Yang and {Zheng}, Jie and {Wu}, Hong and {Liu}, Jifeng},
        title = "{The Mini-SiTian Array: Prospects of Searching for Tidal Disruption Events}",
      journal = {Research in Astronomy and Astrophysics},
         year = 2025,
        month = apr,
       volume = {25},
       number = {4},
          eid = {044011},
        pages = {044011},
          doi = {10.1088/1674-4527/adc78a},
archivePrefix = {arXiv},
       eprint = {2501.09390},
 primaryClass = {astro-ph.HE},
       adsurl = {https://ui.adsabs.harvard.edu/abs/2025RAA....25d4011W}
}

@ARTICLE{wfst2,
       author = {{Lei}, Lei and {Zhu}, Qing-Feng and {Kong}, Xu and {Wang}, Ting-Gui and {Zheng}, Xian-Zhong and {Shi}, Dong-Dong and {Fan}, Lu-Lu and {Liu}, Wei},
        title = "{Limiting Magnitudes of the Wide Field Survey Telescope (WFST)}",
      journal = {Research in Astronomy and Astrophysics},
         year = 2023,
        month = mar,
       volume = {23},
       number = {3},
          eid = {035013},
        pages = {035013},
          doi = {10.1088/1674-4527/acb877},
archivePrefix = {arXiv},
       eprint = {2301.03068},
 primaryClass = {astro-ph.IM},
       adsurl = {https://ui.adsabs.harvard.edu/abs/2023RAA....23c5013L}
}

@ARTICLE{2025JATIS..11b6003L,
       author = {{Layden}, Christopher and {Juneau}, Jill and {Pettersson}, Gustav and {Lourie}, Nathan and {Schneider}, Benjamin and {LaMarr}, Beverly and {Elio Angile}, Francesco and {Farag}, Fadi and {Luo}, Michelle and {Ong}, Zhi Zheng and {Furesz}, Gabor},
        title = "{Characterization of the Teledyne COSMOS camera: a large format CMOS image sensor for astronomy}",
      journal = {Journal of Astronomical Telescopes, Instruments, and Systems},
         year = 2025,
        month = apr,
       volume = {11},
          eid = {026003},
        pages = {026003},
          doi = {10.1117/1.JATIS.11.2.026003},
archivePrefix = {arXiv},
       eprint = {2502.00101},
 primaryClass = {astro-ph.IM},
       adsurl = {https://ui.adsabs.harvard.edu/abs/2025JATIS..11b6003L}
}

@misc{TeledyneCOSMOS,
  author       = {{Teledyne Vision Solutions}},
  title        = {COSMOS sCMOS Camera},
  howpublished = {\url{https://www.teledynevisionsolutions.com/products/cosmos/}},
  year         = {2025},
  note         = {Accessed: 2025-12}
}
\bibliographystyle{aasjournal}

\end{document}